\documentclass[12pt,letterpaper]{article}
\usepackage[letterpaper,margin=1in]{geometry}
\usepackage{microtype}
\usepackage{authblk}
\input{__macros}

\title{Full-Rank Constructions of Explicit Lossless Expanders and Good Small-Set qLTCs}

\author[1]{Yali Aharon Holzman\thanks{The research leading to these results
has received funding from the Israel Science Foundation (grant number 443/22). Email: {\tt yali1@mail.tau.ac.il}.}}
\author[2]{Michael Ilan\thanks{Supported by the European Research Council (ERC) grant number 101201649. \newline
Email: {\tt Michael.ilan@mail.huji.ac.il}.}}
\author[3]{Assaf Reiner\thanks{Supported by the European Research Council (ERC) under the European Union’s Horizon 2020 (N. 882751). Email: {\tt assaf.reiner@mail.huji.ac.il}.}}
\author[1]{Amnon~Ta-Shma\thanks{The research leading to these results
has received funding from the Israel Science Foundation (grant number 443/22). Email: {\tt amnon@tauex.tau.ac.il}.}}
\affil[1]{Department of Computer Science, Tel Aviv University, Israel}
\affil[2]{School of Computer Science and Engineering, The Hebrew University of Jerusalem, Israel}
\affil[3]{Department of Mathematics, The Hebrew University of Jerusalem, Israel}

\date{}

\hypersetup{pdftitle={Full-Rank Constructions of Explicit Lossless Expanders and Good Small-Set qLTCs},
  colorlinks=true, linkcolor=blue!60!black, citecolor=green!40!black,
  urlcolor=blue!60!black}

\begin{document}
\maketitle

\begin{abstract}
We construct explicit two-sided lossless expanders whose binary adjacency matrices have full rank, and use them to obtain explicit asymptotically good quantum LDPC codes whose specified \(X\)- and \(Z\)-check matrices have full row rank and satisfy small-set soundness, or confinement, on a linear range.

Our lossless expanders are a variant of those of Hsieh, Lubotzky, Mohanty, Reiner, and Zhang \cite{HLMRZ25}, with a free action of a finite \(2\)-group. Under such an action, injectivity and surjectivity of the adjacency matrix are detected on a fixed-size matrix indexed by vertex orbits. Adding a sparse factor to the gadget at the middle vertices of a single type then enforces full rank, with a controlled loss in expansion. To obtain the \(2\)-group action, we observe that reducing the generators of the Lubotzky--Phillips--Sarnak Ramanujan graphs modulo powers of \(2\), instead of modulo an odd prime, gives Ramanujan Cayley graphs on finite \(2\)-groups. This construction may be of independent interest.

Plugging two such expanders into the balanced product of Lin and Hsieh \cite{LH22Quantum} gives the quantum codes, since full rank of the input graphs is inherited by the check matrices.

Our construction supplies the code family sought by Hong, Guo, and Lucas \cite{HGL25}. Applying their analysis, within their framework and under their dynamical assumptions, gives quantum memories whose free-energy density is analytic at every positive temperature and whose guaranteed lifetime is \(\exp(\Omega(N))\) at sufficiently low fixed positive temperature, where \(N\) is the blocklength. This improves the previously guaranteed stretched-exponential timescale.

\end{abstract}

\tableofcontents
\clearpage

\section{Introduction}
\label{sec:introduction}

Hong, Guo, and Lucas \cite{HGL25} showed that quantum information can remain protected for long times at low positive temperature even when the free-energy density is analytic at every positive temperature. Their construction brings together two properties of quantum LDPC codes: small-set confinement and independence of the specified sparse checks. Confinement supplies an energy penalty for errors, while independence makes the partition function exactly computable. Their work thus separates the dynamical protection of quantum information from the presence of a thermodynamic phase transition.

Their quantum examples have positive rate, but their distance and confinement range are of order \(\sqrt{N}\), where \(N\) is the number of physical qubits, and their analysis guarantees a stretched-exponential memory timescale. They leave open whether the same combination can be achieved with asymptotically good quantum codes and confinement on a linear range. We answer this coding-theoretic question by constructing an explicit family of asymptotically good binary CSS quantum LDPC codes whose specified \(X\)- and \(Z\)-check matrices have full row rank and satisfy confinement on a linear range. The requirement is simultaneous: independence and confinement hold for the same sparse checks.

The physical consequences follow by applying the analysis of \cite{HGL25} with our improved code parameters, within their framework and under their dynamical assumptions. This gives a guaranteed memory lifetime of \(\exp(\Omega(N))\) at sufficiently low fixed positive temperature, while preserving thermodynamic analyticity. We refer to \cite{HGL25} for the physical interpretation and dynamical analysis. We now explain the coding problem and the challenges in obtaining both properties simultaneously.

\subsection{Small-set local testability and independent checks}

\paragraph{The classical case.}
A binary linear code is specified by a parity-check matrix
\[
    H:\mathbb F_2^n\longrightarrow\mathbb F_2^m,
    \qquad C=\ker H.
\]
The code is low-density parity-check (LDPC) when the row and column
weights of \(H\) are bounded by constants: each check reads constantly
many bits, and each bit participates in constantly many checks.
Full row rank means that the specified checks are linearly independent,
in which case \(\dim C=n-m\).

Suitable random biregular Tanner graphs give asymptotically good
classical LDPC codes. Let \(d\) be the degree of a bit vertex and
\(d'>d\) the degree of a check vertex, so that \(m=(d/d')n\).
For suitable fixed degrees, the random graph has, with high probability,
lossless expansion on a linear range: there are constants
\(\alpha>0\) and \(\varepsilon<1/2\) such that
\[
    |\Gamma(S)|\ge (1-\varepsilon)d|S|
    \qquad\text{for every }S\subseteq[n]\text{ with }|S|\le\alpha n.
\]
At least \((1-2\varepsilon)d|S|\) check vertices then have exactly one
neighbor in \(S\). Each such check is violated by the error supported
on \(S\). Consequently,
\begin{equation}
    |He|\ge\gamma|e|
    \qquad\text{whenever }|e|\le\alpha n,
    \qquad \gamma=(1-2\varepsilon)d.
    \label{eq:classical-confinement}
\end{equation}
In particular, every nonzero codeword has weight greater than
\(\alpha n\).

This also gives small-set local testability. If
\(\operatorname{dist}(y,C)\le\alpha n\), choose a nearest codeword
\(c\in C\) and apply \eqref{eq:classical-confinement} to \(e=y+c\).
Since \(He=Hy\), we obtain
\[
    |Hy|\ge\gamma\,\operatorname{dist}(y,C).
\]
As \(m=\Theta(n)\), a uniformly sampled check rejects \(y\) with
probability at least a constant times its relative distance from
\(C\), throughout this small-set range.

When \(d\) is odd, suitable random biregular ensembles also give
full-row-rank matrices with high probability; see Appendix~A of the
preprint of \cite{HGL25}. Oddness avoids an immediate dependency:
if every column has even weight, the sum of all rows is zero.
Thus asymptotically good classical LDPC codes can simultaneously have
independent sparse checks and small-set local testability.

\paragraph{Testing over the full range.}
To obtain an LTC tester from the specified matrix \(H\), one requires
a constant \(\kappa>0\), independent of \(n\), such that
\[
    |Hy|\ge\kappa\,\operatorname{dist}(y,C)
    \qquad\text{for every }y\in\mathbb F_2^n.
\]
For asymptotically good codes, such a constant-query tester must use
linearly many redundant checks beyond a basis of the dual code,
as shown by Ben-Sasson, Guruswami, Kaufman, Sudan, and
Viderman \cite{BSGKSV10}.

For the full-rank lossless matrices above, the obstruction can be seen
directly. Choose the parameters so that \(\gamma>1\).
Surjectivity of \(H\) gives a word \(y\) whose syndrome has weight one.
By \eqref{eq:classical-confinement}, every vector with this syndrome
has weight greater than \(\alpha n\). Adding a codeword preserves
the syndrome, so
\[
    \operatorname{dist}(y,C)>\alpha n,
    \qquad |Hy|=1.
\]
Hence the specified checks cannot provide a constant-soundness test
over the full range. This obstruction concerns the matrix \(H\);
it does not rule out a different tester using redundant checks
for the same code.

\paragraph{The quantum problem.}
We seek the same combination for quantum codes: positive rate,
linear distance, small-set soundness on a linear range, and independent
sparse checks. A binary CSS code is specified by two matrices
\[
    H_X\in\mathbb F_2^{m_X\times N},
    \qquad
    H_Z\in\mathbb F_2^{m_Z\times N},
    \qquad
    H_XH_Z^T=0.
\]
Their rows specify commuting \(X\)- and \(Z\)-type checks, respectively,
and their columns correspond to qubits. Commutation couples the two
matrices, so the classical random construction cannot simply be
applied independently to the two types of checks.

There is a further distinction: a nonzero error can be a stabilizer
and therefore act trivially on the code. Error size must consequently
be measured after reduction modulo stabilizers. For pure \(X\)- and
\(Z\)-errors, represented by \(x,z\in\mathbb F_2^N\), define
\begin{equation}
    r_X(x)=\min_{a\in\mathbb F_2^{m_X}}|x+H_X^Ta|,
    \qquad
    r_Z(z)=\min_{b\in\mathbb F_2^{m_Z}}|z+H_Z^Tb|.
    \label{eq:reduced-weights}
\end{equation}
We require constants \(\alpha,\gamma>0\) such that
\begin{equation}
    \begin{aligned}
        |H_Zx|&\ge\gamma r_X(x)
        &&\text{whenever }r_X(x)\le\alpha N,\\
        |H_Xz|&\ge\gamma r_Z(z)
        &&\text{whenever }r_Z(z)\le\alpha N.
    \end{aligned}
    \label{eq:quantum-confinement}
\end{equation}
We call these the \emph{small-set soundness} inequalities, with
\emph{soundness coefficient} \(\gamma\) on the range \(\alpha N\). They give
the small-set quantum local testability guarantee used here. In the
physics literature this property is called confinement
\cite{HGL25}. The reductions in
\eqref{eq:reduced-weights} are modulo stabilizers, not modulo the
entire syndrome kernels. A nontrivial logical operator has zero
syndrome but positive reduced weight, so
\eqref{eq:quantum-confinement} also implies quantum minimum distance
greater than \(\alpha N\).

The additional requirement is
\[
    \operatorname{rank}H_X=m_X,
    \qquad
    \operatorname{rank}H_Z=m_Z.
\]
As in the classical case, independence and soundness must hold
for the same specified sparse checks. Deleting dependent checks
preserves the code and sparsity, but can substantially decrease
syndrome weights. Selecting a basis of the check rows therefore
does not by itself establish the desired combination.

\paragraph{Previous quantum constructions.}
The hypergraph-product examples studied by Hong, Guo, and
Lucas \cite{HGL25} already combine positive constant rate,
independent sparse checks, and small-set soundness. Their soundness
guarantee, however, holds only up to reduced weight \(c\sqrt{N}\),
for a constant \(c>0\), and their distance is \(O(\sqrt{N})\).
The problem is to achieve linear distance and small-set soundness on a
linear range while retaining independence of the specified checks.

\paragraph{Our results.}
Our first result is a family of explicit lossless expanders with
full-rank adjacency matrices. A \((\Delta_L,\Delta_R)\)-biregular
bipartite graph on \(L\sqcup R\) is \emph{two-sided
\((c,\varepsilon)\)-lossless} if every set \(S\) of at most \(c|L|\)
left vertices has at least \((1-\varepsilon)\Delta_L|S|\) neighbors, and
symmetrically for sets of right vertices.

\begin{theorem}[Full-rank lossless expanders]
\label{thm:intro-full-rank-lossless}
For every integer \(t\ge1\), every \(\varepsilon>0\), and every \(a_{min} \geq 1\), there are a positive integer \(a \geq a_{min}\), a constant \(c>0\), and an explicit
infinite family of \((a,ta)\)-biregular bipartite graphs on
\(L\sqcup R\) with \(|L|=t|R|\), such that:
\begin{enumerate}
    \item
    each graph is two-sided \((c,\varepsilon)\)-lossless;

    \item
    its binary adjacency map \(\mathbb F_2^L\to\mathbb F_2^R\) has full
    row rank, and is invertible when \(t=1\);

    \item
    it admits a free action of a finite \(2\)-group with a bounded
    number of orbits.
\end{enumerate}
The neighbors of every vertex can be computed in time polynomial in
the logarithm of the number of vertices.
\end{theorem}

Plugging two such graphs into the Lin--Hsieh balanced product gives
quantum LDPC codes with the combination described above. The rate can
be made arbitrarily close to one, and the soundness coefficient can be
prescribed in advance.

\begin{theorem}[Independent checks and small-set quantum local testability]
\label{thm:intro-independent-checks}
Let \(t_1,t_2\ge2\) be integers and let \(\gamma_0>0\).
There are constants \(\alpha>0\) and \(w\in\mathbb Z_{>0}\), and an
explicit infinite family of binary CSS codes with blocklengths
\(N\to\infty\), whose specified parity-check matrices \(H_X,H_Z\)
satisfy the following properties:
\begin{enumerate}
    \item
    \(H_XH_Z^T=0\), both matrices have full row rank, and every row
    and column of either matrix has weight at most \(w\).
    Each qubit participates in at most \(2w\) checks in total.

    \item
    The number \(K\) of encoded qubits satisfies
    \[
        \frac{K}{N}
        =
        \frac{(t_1-1)(t_2-1)}{t_1t_2+1}.
    \]

    \item
    The small-set soundness inequalities \eqref{eq:quantum-confinement}
    hold with \(\gamma=\gamma_0\).
    In particular, the quantum minimum distance is greater than
    \(\alpha N\).
\end{enumerate}
Every row and column of \(H_X,H_Z\) can be enumerated in time
polynomial in \(\log N\), after finite precomputation depending
only on \(t_1,t_2,\gamma_0\).
\end{theorem}

Taking \(t_1=t_2=4\) gives rate \(9/17\), while taking both integers
large makes the rate approach one. The constants \(\alpha\) and
\(w\) may depend on the chosen rate and soundness coefficient.
The theorem establishes small-set quantum local testability on a
linear range; it does not assert quantum local testability over
the full range.

\subsection{Techniques}
\label{sec:intro-techniques}

The construction has two rank reductions. First, we reduce independence
of the quantum checks to rank conditions on the input graphs of the
Lin--Hsieh balanced product. Second, we show that, in a suitable variant
of the HLMRZ construction, these conditions reduce to a rank test on a
fixed finite gadget. The second reduction requires a free action of a
finite \(2\)-group. To supply this symmetry together with the required
expansion, we construct LPS Ramanujan cubical complexes on a common
family of finite \(2\)-groups.

\paragraph{Rank inheritance in the Lin--Hsieh construction.}
Lin and Hsieh use the balanced product to construct a two-dimensional
cubical object from two bipartite graphs with a common free group
action \cite{BE21,LH22Quantum}. Directional losslessness is inherited
from the input graphs, and their small-set estimate gives small-set
soundness for a suitable choice of parameters (see also \cite{CLRTT26} for a
higher-dimensional version). Our observation is that appropriate rank
conditions are inherited as well.

Write the two input adjacency maps as
\[
    A:\mathbb F_2^{A_0}\longrightarrow\mathbb F_2^{A_1},
    \qquad
    B:\mathbb F_2^{B_0}\longrightarrow\mathbb F_2^{B_1}.
\]
Their balanced product has vertex blocks
\[
    V_{ij}=(A_i\times B_j)/G,
    \qquad i,j\in\{0,1\},
\]
and the three-term complex
\[
    \mathbb F_2^{V_{00}}
    \xrightarrow{\delta_0}
    \mathbb F_2^{V_{10}}\oplus\mathbb F_2^{V_{01}}
    \xrightarrow{\delta_1}
    \mathbb F_2^{V_{11}}.
\]
The associated CSS code has check matrices
\[
    H_X=\delta_0^T,
    \qquad
    H_Z=\delta_1.
\]

Freeness of the group action makes each directional map a direct sum
of copies of the corresponding input map. Consequently, injectivity
of \(A\) implies injectivity of \(\delta_0\), and surjectivity of \(B\)
implies surjectivity of \(\delta_1\). These are precisely the conditions
that make \(H_X\) and \(H_Z\) have full row rank. Section~2 gives the
argument. The remaining task is therefore to construct two-sided
lossless bipartite graphs of suitable opposite imbalances, with
full-rank adjacency maps and a common free group action.

\paragraph{Detecting rank on vertex orbits.}
The construction of Hsieh, Lubotzky, Mohanty, Reiner, and Zhang
\cite{HLMRZ25}, which we call HLMRZ, builds two-sided lossless graphs
from cubical incidence graphs and a fixed finite gadget. Their construction retains a free group action,
but does not supply the rank guarantee needed here. Our modification
uses a special feature of finite \(2\)-groups acting in characteristic
two.

If a finite \(2\)-group acts linearly on a nonzero binary vector space,
it fixes a nonzero vector. Indeed, the number of nonzero vectors is odd,
whereas every orbit has power-of-two size, so at least one orbit must
be a singleton. Applying this observation to the kernel of an
equivariant adjacency map shows that every nonzero kernel contains a
nonzero vector constant on each vertex orbit. Thus injectivity is
detected by the restriction to orbit-constant vectors.

We call the matrix of this restriction the \emph{orbit matrix}.
Since the action is free, the orbit matrix of the transpose is the
transpose of the orbit matrix. Applying the same kernel argument
to the transpose therefore detects surjectivity as well. In our
construction, the numbers of vertex orbits are fixed while the
graphs grow, so this reduces the rank problem to a matrix of fixed
size.

\paragraph{Changing the gadget at one middle type.}
In the HLMRZ line product, the vertex orbits on each side are indexed
by cube signatures. Let \(H\) be the set of middle types, with
\(h=|H|\) even, and let \(C_x\) be the gadget matrix used at middle
vertices of type \(x\in H\). Our choice of input complexes ensures
that a left cube and a right cube share at most one middle vertex.
This distinct-witness property lets us count the edge contributions
from the different middle types separately: the orbit matrix of
the line product is
\[
    \sum_{x\in H} C_x.
\]
In the unmodified construction, the same gadget \(C\) is used at
every type, so this matrix is \(hC=0\). We break this cancellation
by changing the gadget at one middle type.

Suppose the left signature set is \(t\) times as large as the right
one, for an integer \(t\ge1\). For the noncomplete biregular gadget
used here, a matching argument gives a matrix \(F\), supported on
gadget nonedges, with column weight one and row weight \(t\).
Its rows have disjoint nonempty supports, so \(F\) has full row rank.
We replace \(C\) by \(C+F\) at every middle vertex of type \(0\),
leaving the other types unchanged. The orbit matrix becomes
\[
    hC+F=F,
\]
and the \(2\)-group rank criterion gives full row rank of the entire
adjacency matrix. Reversing the bipartition supplies the injective
map needed for the other input.

The modification preserves the group action because it depends only
on the middle type. The nonedge support of \(F\), together with the
distinct-witness property, ensures that all inserted edges are new.
If the original line product has degrees \((a_0,ta_0)\), the modified
degrees are
\[
    (a_0+1,\;t(a_0+1)).
\]
Since edges are only added, every original neighborhood is preserved.
Thus the same lossless range remains valid, and an original loss
\(\varepsilon_0\) increases to at most
\[
    \varepsilon_+
    =
    \frac{\varepsilon_0a_0+1}{a_0+1}
    =
    \varepsilon_0+\frac{1-\varepsilon_0}{a_0+1}.
\]
Section~3 proves the modification and this loss bound.

\paragraph{LPS Ramanujan graphs on $2$-groups.}
The remaining issue is to construct a cubical complex that can serve as 
the input to the HLMRZ construction and that admits a free action of a 
finite $2$-group with a bounded number of orbits. In the HLMRZ construction
the underlying group is $\operatorname{PSL}_2(\mathbb{F}_q)$. It comes from
the graphs in the directions of the complex, which are the LPS Ramanujan
graphs~\cite{LPS88}. Thus, a natural approach is to carry out an
analogue of the LPS construction of Ramanujan graphs in which the
underlying group is a $2$-group. This is the program we carry out in \cref{sec:2group-cubical}. 

At first glance, constructing bounded-degree expanders from finite
$p$-groups might seem to contradict known non-expansion results. In
particular, Lubotzky and Weiss~\cite{LW93} proved that finite solvable
groups of \emph{bounded derived length} cannot yield bounded-degree
expanders. Since finite $p$-groups are nilpotent, and thus solvable, one
might suspect that they cannot expand. However, the obstruction
of~\cite{LW93} applies only when the derived length is uniformly
bounded. In our setting the derived length tends to infinity, and expansion is entirely possible. In
fact, Lubotzky and Weiss themselves demonstrated this with an infinite
family of expanders on finite $p$-groups. We briefly review their
example, as it clarifies some steps in our approach.

Let $\Gamma=\operatorname{SL}_n(\mathbb{Z})$, and let $\Gamma(N)$ be the
principal congruence subgroup of level $N$, namely the kernel of
reduction modulo $N$. Fix a prime $p$. For $j\ge1$ we have
$\Gamma(p^{j+1})\trianglelefteq\Gamma(p^j)$, and
$\Gamma(p^j)/\Gamma(p^{j+1})$ is an abelian $p$-group of
order $p^{n^2-1}$. This follows by a direct computation or, more
conceptually, from the isomorphism onto the additive group of the Lie
algebra
\[
 \Gamma(p^j)/\Gamma(p^{j+1})\xrightarrow{\ \sim\ }
 \mathfrak{sl}_n(\mathbb{F}_p),
 \qquad I+p^jX\longmapsto X\bmod p.
\]
Thus $\Gamma(p)/\Gamma(p^{m})$ is a finite $p$-group, of order
\[
 |\Gamma(p)/\Gamma(p^{m})|
 =\prod_{j=1}^{m-1}|\Gamma(p^{j})/\Gamma(p^{j+1})|
 =p^{(m-1)(n^{2}-1)}.
\]
These quotients form expanders with respect to suitable bounded-size
generating sets. For instance, for $n\ge3$ this follows from Margulis'
argument~\cite{Mar73}, since $\Gamma$ has property~(T) and $\Gamma(p)$
is a finite-index subgroup.

Now back to our setting. One can verify by similar reasoning that if
$\Gamma$ is an arithmetic subgroup of an algebraic group $G$ that is
smooth at $p$, then $\Gamma(p)/\Gamma(p^m)$ is a $p$-group of order at
most $p^{(m-1)\dim G}$, with equality when $\Gamma$ is dense in
$G(\mathbb Z_p)$.

Thus, to carry out an analogue of the LPS Ramanujan graphs, we recall
that their construction uses the group $\Gamma=G'(\mathbb Z[1/p])$ for a
prime $p\equiv1\pmod4$, where $G'=D^{\times}/Z(D^{\times})$ and
$D=\bigl(\frac{-1,-1}{\mathbb{Q}}\bigr)$ is the algebra of Hamilton
quaternions over $\mathbb Q$. The integral structure
is given by the Lipschitz quaternions: $\Gamma$ is the group of units of
$\Lips(\mathbb Z[1/p])$, namely the quaternions whose norm is a power
of $p$, modulo the scalars $\pm p^{z}$.
There are $8(p+1)$ quaternions in $\Lips(\mathbb Z)$ of norm $p$. Each
of them has exactly one associate under the units
$\pm1,\pm\qi,\pm\qj,\pm\qk$ with positive odd real part, and these $p+1$
representatives are denoted $S_p$. They generate the principal
congruence subgroup $\Gamma(2)$, which is a free group on $(p+1)/2$
generators. This group acts simply transitively on the vertices of the
$(p+1)$-regular tree $T_{p+1}$, inducing an isomorphism:
\[
 T_{p+1}\cong\operatorname{Cay}(\Gamma(2),S_p).
\]
In the LPS setting one now takes the quotient modulo an odd prime
$q\ne p$, namely by $\Gamma(2q)\trianglelefteq\Gamma(2)$, to obtain a
finite Cayley graph on $\Gamma(2)/\Gamma(2q)$. LPS identify this group
with $\operatorname{PSL}_2(\mathbb{F}_q)$ or
$\operatorname{PGL}_2(\mathbb{F}_q)$, according to whether $p$ is a
quadratic residue modulo $q$. The Ramanujan property relies on a deep
input, which is Deligne's proof of the Ramanujan--Petersson conjecture. In the
form of \cite[Theorem~7.3.1]{Lub94}, it gives the Ramanujan bound for
$\Gamma(N)\backslash T_{p+1}$ at every level $N$ prime to $p$.

Our setting requires the groups to be $2$-groups, and the discussion
above suggests a natural way to achieve this. Since
$\Gamma(2^m)\trianglelefteq\Gamma(2)$, one can take the LPS generators
and reduce them modulo $2^m$. By the reasoning above,
$\Gamma(2)/\Gamma(2^m)$ is a $2$-group; it embeds in a $2$-group of
order $2^{(m-1)\dim G'}=2^{3(m-1)}$. Since
$2^m$ is prime to $p$, the LPS machinery gives the Ramanujan bound for
the resulting Cayley graphs. 

As in the LPS dichotomy between $\operatorname{PSL}_2$ and
$\operatorname{PGL}_2$, the quotient depends on whether $p$ is a square
modulo $2^m$, which happens exactly when $p\equiv1\pmod8$. For such $p$,
strong approximation shows that $\Gamma(2)/\Gamma(2^m)$ is an index-two
subgroup $G_m$ of order $2^{3m-4}$, which does not depend on $p$. Hence the LPS graphs of several
primes $p_1,\ldots,p_\ell\equiv1\pmod8$ live on the same $2$-group
$G_m$. The cubical structure needed for the HLMRZ construction then
follows as in their original construction, from the unique
factorization of odd integral quaternions due to
Dickson~\cite{Dic22}.

We point out that, fortunately, the LPS construction is very well
adapted to the $2$-group setting we need. The LPS generators already lie in
$\Gamma(2)$, the top of the tower
$\Gamma(2)\supseteq\Gamma(4)\supseteq\cdots\supseteq\Gamma(2^m)$, so
their reductions modulo $2^m$ generate $2$-groups. This is not the case
for an odd prime $q\ne p$. The quotients $\Gamma(2)/\Gamma(2q^m)$ are
not $q$-groups, since they map onto $\Gamma(2)/\Gamma(2q)$, which is
$\operatorname{PSL}_2(\mathbb F_q)$ or
$\operatorname{PGL}_2(\mathbb F_q)$, as the case $m=1$ above already
shows.

\subsection{Concluding remarks}

Our construction supplies asymptotically good quantum LDPC codes
with independent sparse checks and small-set soundness on a linear
range.
Applying the analysis of Hong, Guo, and Lucas \cite{HGL25}, within
their framework and under their dynamical assumptions, improves
the guaranteed memory lifetime from a stretched-exponential timescale
to \(\exp(\Omega(N))\) at sufficiently low fixed positive temperature.
The free-energy density remains analytic at every positive temperature.

The proof reduces independence of the quantum checks to rank
conditions on two lossless graphs, and then reduces these conditions
to a rank test on a fixed finite gadget. The second reduction relies
on the acting group being a finite \(2\)-group. The original HLMRZ
construction \cite{HLMRZ25} uses the groups
\(\mathrm{PSL}_2(\mathbb F_q)\), so our rank-detection criterion does
not apply. We wonder whether a different argument can enforce the
required rank conditions through a modification of a fixed finite
gadget, while retaining the original group family and lossless
expansion.

\section{The Lin--Hsieh construction preserves full rank}
\label{app:LH}

Lin and Hsieh use the balanced product to build a two-dimensional
cubical complex from two bipartite graphs. A key feature of their
construction is that directional lossless expansion of the resulting
two-dimensional complex is inherited from the lossless expansion of the
two input graphs. In this section, we observe that the same inheritance
principle holds for rank: suitable full-rank properties of the adjacency
matrices of the two input graphs imply the corresponding full-rank
properties of the boundary maps, and hence of the parity-check matrices,
of the resulting complex. We now give the details.

\subsection{The balanced product}
\label{app:LH-balanced-product}

Let
\[
\Gamma_A=(A_0,A_1,E_A)
\qquad\text{and}\qquad
\Gamma_B=(B_0,B_1,E_B)
\]
be two bipartite graphs, and denote their adjacency matrices by $A$ and
$B$. Suppose that the two graphs admit a common free $G$-action. Their
balanced product has four blocks of vertices
\[
V_{ij}=(A_i\times B_j)/G,
\qquad i,j\in\{0,1\},
\]
where the quotient is with respect to the diagonal $G$-action
\[
(v,w)\sim(gv,gw),\qquad g\in G.
\]
The edges are quotiented by the same action:
\[
((v,w),(v',w'))
   \sim
((gv,gw),(gv',gw')),
\qquad g\in G.
\]

Denote by
\[
S_j:\mathbb F_2^{V_{0j}}\longrightarrow
     \mathbb F_2^{V_{1j}},
\qquad
T_i:\mathbb F_2^{V_{i0}}\longrightarrow
     \mathbb F_2^{V_{i1}}
\]
the two families of directional adjacency maps. They form the diagram
\[
\begin{tikzcd}[column sep=large,row sep=large]
\mathbb F_2^{V_{00}}
    \arrow[r,"S_0"]
    \arrow[d,"T_0"']
&
\mathbb F_2^{V_{10}}
    \arrow[d,"T_1"]
\\
\mathbb F_2^{V_{01}}
    \arrow[r,"S_1"']
&
\mathbb F_2^{V_{11}} .
\end{tikzcd}
\]

The following directional-copy property is the basic structural fact that
we need.

\begin{lemma}[{\cite[Lemma 26]{LH22Quantum}}]
\label{lem:LH-directional-copies}
The subgraph between $V_{00}$ and $V_{10}$ defined by $S_0$ is a
disjoint union of $|B_0/G|$ copies of $\Gamma_A$. Equivalently, after
suitable orderings of the bases,
\[
S_0\cong\bigoplus_{B_0/G} A.
\]
The analogous statement holds for the other three directional maps:
$S_1$ is a direct sum of copies of $A$, while $T_0$ and $T_1$ are
direct sums of copies of $B$.
\end{lemma}

\begin{proof}
Fix a $G$-orbit $O\subseteq B_0$ and a representative $b\in O$.
Since the $G$-action on $B_0$ is free, every orbit in
$(A_i\times O)/G$ has a unique representative of the form $[a,b]$.
Thus
\[
a\longmapsto[a,b]
\]
identifies the subgraph induced on $(A_0\times O)/G$ and
$(A_1\times O)/G$ with $\Gamma_A$. Different $G$-orbits in $B_0$ give
disjoint components. Hence the graph defined by $S_0$ is the disjoint
union of $|B_0/G|$ copies of $\Gamma_A$, and consequently
\[
S_0\cong\bigoplus_{B_0/G} A.
\]
The other three statements follow identically by fixing an orbit in the
unused coordinate.
\end{proof}

The balanced-product squares imply
\[
T_1S_0=S_1T_0.
\]
Indeed, taking one step in each of the two graph directions is independent
of the order of the two steps. Consequently the balanced product defines
the three-term complex
\begin{equation}
\label{eq:LH-complex}
C^0=\mathbb F_2^{V_{00}}
\xrightarrow{\delta_0}
C^1=\mathbb F_2^{V_{10}}\oplus\mathbb F_2^{V_{01}}
\xrightarrow{\delta_1}
C^2=\mathbb F_2^{V_{11}},
\end{equation}
where
\begin{equation}
\label{eq:LH-boundary-maps}
\delta_0=
\begin{pmatrix}
S_0\\
T_0
\end{pmatrix},
\qquad
\delta_1=
\begin{pmatrix}
T_1&S_1
\end{pmatrix}.
\end{equation}
The identity $T_1S_0=S_1T_0$ gives
\[
\delta_1\delta_0=0
\]
over $\mathbb F_2$.

The associated CSS code places the qubits on
$V_{10}\sqcup V_{01}$, the $X$-checks on $V_{00}$, and the
$Z$-checks on $V_{11}$. Its parity-check matrices are
\begin{equation}
\label{eq:LH-checks}
H_X=\delta_0^T=
\begin{pmatrix}
S_0^T&T_0^T
\end{pmatrix},
\qquad
H_Z=\delta_1=
\begin{pmatrix}
T_1&S_1
\end{pmatrix}.
\end{equation}

\subsection{Rank inheritance}
\label{app:LH-rank}

We now show that the rank properties of the two input graphs are inherited
by the two parity-check matrices of the balanced product. The two ends of
the chain complex require opposite rank conditions: independence of the
$X$-checks amounts to injectivity of $\delta_0$, whereas independence of
the $Z$-checks amounts to surjectivity of $\delta_1$. Accordingly, we
require the adjacency map of the first input graph to be injective and that
of the second to be surjective.

\begin{theorem}[Rank inheritance]
\label{thm:LH-rank-inheritance}
Let
\[
\Gamma_A=(A_0,A_1,E_A)
\qquad\text{and}\qquad
\Gamma_B=(B_0,B_1,E_B)
\]
be two bipartite graphs with a common free $G$-action, and let
\[
A:\mathbb F_2^{A_0}\longrightarrow\mathbb F_2^{A_1},
\qquad
B:\mathbb F_2^{B_0}\longrightarrow\mathbb F_2^{B_1}
\]
be their adjacency maps. If $A$ is injective and $B$ is surjective,
then the two parity-check matrices $H_X$ and $H_Z$ of the Lin--Hsieh
balanced-product code \cite{LH22Quantum} have full row rank.
\end{theorem}

\begin{proof}
Suppose first that $a\in\mathbb F_2^{V_{00}}$ is a dependency among
the rows of $H_X$. By~\eqref{eq:LH-checks},
\[
0=H_X^Ta
  =\delta_0a
  =
  \begin{pmatrix}
  S_0a\\
  T_0a
  \end{pmatrix},
\]
and hence $S_0a=0$. By
\cref{lem:LH-directional-copies}, $S_0$ is a direct sum of copies
of $A$. Since $A$ is injective, so is $S_0$, and therefore $a=0$.
Thus $H_X$ has full row rank.

Similarly, suppose that $b\in\mathbb F_2^{V_{11}}$ is a dependency
among the rows of $H_Z$. Then
\[
0=H_Z^Tb
  =\delta_1^Tb
  =
  \begin{pmatrix}
  T_1^Tb\\
  S_1^Tb
  \end{pmatrix},
\]
and hence $T_1^Tb=0$. By
\cref{lem:LH-directional-copies}, $T_1$ is a direct sum of copies
of $B$. Since $B$ is surjective, $B^T$ is injective, and therefore
$T_1^T$ is injective. Hence $b=0$, and $H_Z$ has full row rank.
\end{proof}
\section{A full-rank modification of HLMRZ}
\label{sec:full-rank-modification}

HLMRZ construct two-sided lossless graphs as line products of two cubical
incidence graphs, recalled in \cref{sec:tripartite-line-product}.
Their graphs admit a free action of the underlying group
(\cite[Proposition~A.3]{HLMRZ25}). For our constructions, we need the
binary adjacency matrix to have full row rank as well. We show that,
when the acting group is a finite $2$-group,
this additional requirement can be enforced by adding a sparse
complementary factor at one middle type. The modification preserves the
vertex sets, the free action, and the lossless range, with a controlled
increase in loss. The key is that injectivity and surjectivity of an
equivariant binary map can then be detected on a fixed-size matrix indexed
by vertex orbits.

\begin{assumption}[Free $2$-group action]
\label{ass:rank-free-action}
The acting group $G$ is a finite $2$-group. Its actions on $L$ and $R$
are free and preserve the two vertex classes. In a family, the numbers of
vertex orbits and all local parameters are fixed before $|G|$ grows.
\end{assumption}

This is an additional algebraic requirement, not an expansion assumption.
The groups $\mathrm{PSL}_2(\mathbb F_q)$ of the original HLMRZ construction do not
satisfy it. The required family will be supplied by
the $2$-groups of \cref{sec:2group-cubical}, through
\cref{thm:2group-lossless}. Here we
prove the modification theorem assuming this symmetry.

\subsection{The HLMRZ construction}
\label{sec:tripartite-line-product}

We recall the definitions from HLMRZ that we need, in the notation of
this paper.

\begin{definition}
\label{def:two-sided-lossless}
A $(\Delta_L,\Delta_R)$-biregular bipartite graph on $L\sqcup R$ is
\emph{two-sided $(\alpha,\varepsilon)$-lossless} if
\[
 \begin{aligned}
 |N(S)|&\ge(1-\varepsilon)\Delta_L|S|
 &&(S\subseteq L,\ |S|\le\alpha|L|),\\
 |N(T)|&\ge(1-\varepsilon)\Delta_R|T|
 &&(T\subseteq R,\ |T|\le\alpha|R|).
 \end{aligned}
\]
\end{definition}

\paragraph{Cubes and signatures.}
Let $G$ be a finite group. Inverse-closed subsets
$A_1,\ldots,A_h\subseteq G$ are \emph{cubical generating sets} if
\[
 A_iA_j=A_jA_i\quad(i\ne j),\qquad
 |A_1\cdots A_h|=\prod_{i=1}^h|A_i|.
\]
They define the decorated Cayley
cubical complex $Y=\operatorname{Cay}(G;A_1,\ldots,A_h)$ on the vertex
set $G\times\mathbb F_2^h$. Its cubes are the collections
$f=\{(f_x,x):x\in\mathbb F_2^h\}$ with $f_x^{-1}f_{x+e_i}\in A_i$ for
all $x\in\mathbb F_2^h$ and $i\in[h]$. The \emph{signature} of $f$ is the
labeling
\[
 (x,i)\longmapsto f_x^{-1}f_{x+e_i}\in A_i
\]
of the directed edges of the cube. Left translation $(gf)_x=gf_x$
preserves signatures. For a signature $\sigma$, put
$\phi_\sigma(x)=f_0^{-1}f_x$, where $f$ is any cube of signature
$\sigma$. This is the product of the labels along any path from $0$ to
$x$, so it depends only on $\sigma$. Thus $\phi_\sigma$ is the unique cube
of signature $\sigma$ with $\phi_\sigma(0)=1$, and every cube of
signature $\sigma$ is $g\phi_\sigma$ for a unique $g\in G$. We write this
cube as $(g,\sigma)$.

\paragraph{Incidence graphs.}
Let $h$ be a power of two, and let $H\subseteq\mathbb F_2^h$ be the
Hadamard code. It consists of $h$ vertex types, contains $0$, and any two
distinct types have Hamming distance $h/2$.
For a nonempty set $S$ of signatures, the incidence graph $I(Y,S)$ is
the bipartite graph between the cubes $G\times S$ and the middle
vertices $M=G\times H$, in which a cube is adjacent to its vertices of
types in $H$. This is the coded cubical incidence graph of HLMRZ, restricted to the signatures in $S$. Thus the cube $(g,\sigma)$ is adjacent to
$(g\phi_\sigma(x),x)$ for $x\in H$. A middle vertex $u=(g,x)$ has
exactly one neighbor of each signature $\sigma\in S$, namely
\begin{equation}
\label{eq:incidence-chart}
 \operatorname{Nbr}_u(\sigma)=\bigl(g\phi_\sigma(x)^{-1},\sigma\bigr).
\end{equation}
Left translation by $G$ acts freely on both sides, and
$\operatorname{Nbr}_{ku}(\sigma)=k\operatorname{Nbr}_u(\sigma)$ for
$k\in G$.

\paragraph{The line product.}
Let $Y_L$ and $Y_R$ be two such complexes of the same dimension $h$ on
the same group $G$, and let $S_L,S_R$ be nonempty sets of signatures of
$Y_L,Y_R$, of sizes $D_L,D_R$. The incidence graphs $I_L=I(Y_L,S_L)$ and
$I_R=I(Y_R,S_R)$ have the vertex sets
\[
 L=G\times S_L,\qquad R=G\times S_R,\qquad M=G\times H.
\]
Write $\operatorname{LNbr}_u$ and $\operatorname{RNbr}_u$ for their
neighbor maps~\eqref{eq:incidence-chart}, defined with the normalized cubes
$\phi^L_\sigma$ and $\phi^R_\tau$. A \emph{gadget} is a simple
$(d_L,d_R)$-biregular bipartite graph on $S_L\sqcup S_R$. Let
$C\in\mathbb F_2^{S_R\times S_L}$ be its adjacency matrix. The line
product $Z_0=Z(I_L,I_R;C)$ is the bipartite graph on $L\sqcup R$ with an
edge
\[
 \operatorname{LNbr}_u(\sigma)\sim\operatorname{RNbr}_u(\tau)
\]
for every $u\in M$ and every gadget edge $(\sigma,\tau)$. We call $u$ a
\emph{witness} of this edge. Every cube has one
middle neighbor of each type, and receives $d_L$ or $d_R$ gadget edges at
each of them. The neighbor maps are equivariant, so left translation by
$G$ acts freely on $Z_0$. The line product is a form of the derandomized
squaring of Rozenman and Vadhan \cite{RV05}. It keeps only the two-step
paths from $L$ through $M$ to $R$ whose two labels form a gadget edge.

We will require the \emph{distinct-witness condition}
\begin{equation}
\label{eq:distinct-witness}
 |N_{I_L}(\ell)\cap N_{I_R}(r)|\le1\qquad(\ell\in L,\ r\in R).
\end{equation}
It says that a left cube and a right cube share at most one middle
vertex. Under~\eqref{eq:distinct-witness}, every edge of $Z_0$ has a
unique witness. Since the gadget is simple, $Z_0$ is then simple and
$(hd_L,hd_R)$-biregular. HLMRZ do not impose this condition, since
their losslessness argument does not need it. We use it to make $Z_0$
simple and to compute its orbit matrix
(\cref{lem:line-product-orbit-matrix}).

The main theorem of HLMRZ shows that suitable line products of this form
are lossless.

\begin{theorem}[{\cite[Theorem~2.2]{HLMRZ25}}]
\label{thm:HLMRZ}
For every $\varepsilon\in(0,1)$ and $\beta\in(0,1]$ there are a power of
two $h$ and an integer $d_0$ such that the following holds for all
integers $d_L,d_R\ge d_0$ with $\beta\le d_L/d_R\le\beta+\varepsilon$.
There is $c>0$ such that, for infinitely many primes $q$, there are two
cubical complexes of dimension $h$ on $\operatorname{PSL}_2(\mathbb F_q)$,
signature sets, and a gadget of degrees $(d_L,d_R)$ whose line product
$Z_0$ is two-sided $(c,\varepsilon)$-lossless.
\end{theorem}

\subsection{Equivariant maps and rank detection}
\label{sec:equivariant-rank-detection}

Let $\Gamma$ be a bipartite graph on $L\sqcup R$ on which $G$ acts freely,
and let $A\colon V\to W$ be its adjacency map, where
$V=\mathbb F_2^L$ and $W=\mathbb F_2^R$. The action permutes
coordinates, and since it preserves edges, $A$ is \emph{equivariant}:
$A(gv)=gA(v)$ for $g\in G$ and $v\in V$. In particular, $A$ maps the
subspace $V^G$ of $G$-invariant vectors into $W^G$. A vector is invariant
exactly when it is constant on each orbit, so $V^G$ is identified with
$\mathbb F_2^{L/G}$, and $W^G$ with $\mathbb F_2^{R/G}$.

Let $[A]\colon\mathbb F_2^{L/G}\to\mathbb F_2^{R/G}$ be the adjacency map of
the quotient graph $\Gamma/G$, whose vertices are the vertex orbits and
whose edges are the edge orbits. We call $[A]$ the \emph{orbit matrix}
of $A$. Since $G$ acts freely, $\Gamma\to\Gamma/G$ is a covering map. Indeed, every edge of $\Gamma/G$
at the orbit of a vertex $r$ lifts to an edge at $r$, and freeness
guarantees that distinct edges at $r$ lie in distinct edge orbits.
Hence $[A]$ is identified with the restriction of $A$ to $V^G$. The same argument applies
to the graph read from $R$ to $L$, whose quotient has adjacency map
$[A^T]=[A]^T$. Altogether,
\begin{equation}
\label{eq:orbit-restriction}
 [A]=A|_{V^G},\qquad [A^T]=[A]^T=A^T|_{W^G}.
\end{equation}
Note that the size of $[A]$ is fixed in a
family satisfying \cref{ass:rank-free-action}, although the
dimensions of $A$ grow with $|G|$.

The next lemma shows that injectivity and surjectivity of $ A $ are equivalent to those properties of $[ A ]$. It is the only place where we use that $G$ is a $2$-group,
through the standard fact that a finite $p$-group acting linearly on a
nonzero finite-dimensional vector space over $\mathbb F_p$ fixes a
nonzero vector.

\begin{lemma}
\label{lem:orbit-rank-detection}
Under \cref{ass:rank-free-action}, $A$ is injective if and
only if $[A]$ is injective, and $A$ is surjective if and only if $[A]$
is surjective.
\end{lemma}

\begin{proof}
Since $\ker[A]=(\ker A)^G$, if $A$
is injective so is $[A]$. Conversely, suppose that $\ker A\ne0$. Then
$G$ acts on the $2^{\dim\ker A}-1$ nonzero vectors of $\ker A$. Each orbit
has power-of-two size and their total is odd, so some orbit has size $1$.
This gives a nonzero vector in $(\ker A)^G=\ker[A]$. Hence $A$ is
injective if and only if $[A]$ is.

For surjectivity, recall that a linear map of finite-dimensional spaces
is surjective if and only if its transpose is injective. The map $A^T$
is the adjacency map of $\Gamma$ read from $R$ to $L$, on which $G$ still
acts freely. So the injectivity part applies to it, and
its orbit matrix is $[A]^T$ by~\eqref{eq:orbit-restriction}. Thus
\[
 A\text{ is surjective}
 \Leftrightarrow A^T\text{ is injective}
 \Leftrightarrow [A]^T\text{ is injective}
 \Leftrightarrow [A]\text{ is surjective}.
\]
\end{proof}

\subsection{The orbit matrix of the line product}
\label{sec:line-product-orbit-matrix}

Keep the notation of \cref{sec:tripartite-line-product}, and
assume the distinct-witness condition~\eqref{eq:distinct-witness}. We now
allow the gadget to depend on the middle type. For each $x\in H$, choose
a binary matrix $C_x\in\mathbb F_2^{S_R\times S_L}$, and use it at every
middle vertex of type $x$. Let
$A_{\mathcal C}\colon\mathbb F_2^L\to\mathbb F_2^R$ be the adjacency map
of the resulting line product, where $\mathcal C=(C_x)_{x\in H}$. When
all $C_x$ equal $C$, this is the original line product $Z_0$, and we
write $A_0$ for its adjacency map.

Left translation preserves the middle types, and the neighbor maps are
equivariant. Hence $G$ still acts freely on the line product. Its vertex
orbits are the sets $G\times\{\sigma\}$ for $\sigma\in S_L$ and
$G\times\{\tau\}$ for $\tau\in S_R$. We index them by the signatures, so
that the orbit matrix is a map
\[
 [A_{\mathcal C}]\colon\mathbb F_2^{S_L}\longrightarrow\mathbb F_2^{S_R}.
\]
Its size does not depend on $|G|$. The next lemma computes it.

\begin{lemma}
\label{lem:line-product-orbit-matrix}
Under the distinct-witness condition~\eqref{eq:distinct-witness},
\[
 [A_{\mathcal C}]=\sum_{x\in H}C_x.
\]
More precisely, the edges witnessed at middle vertices of type $x$ have
orbit matrix $C_x$.
\end{lemma}

\begin{proof}
By \cref{sec:equivariant-rank-detection}, $[A_{\mathcal C}]$ is
the adjacency map of the quotient graph over $\mathbb F_2$. Thus its
$(\tau,\sigma)$ entry is the number of edge orbits between
$G\times\{\sigma\}$ and $G\times\{\tau\}$, modulo $2$. We count these
orbits.

Every edge of the line product has the form
$\operatorname{LNbr}_u(\sigma)\sim\operatorname{RNbr}_u(\tau)$, where
$u$ is a middle vertex of some type $x$ and $(\sigma,\tau)$ is an edge of
the gadget $C_x$. Here $\sigma$ and $\tau$ are the signatures of the two
endpoints, and $u$ is the witness of the edge, which is unique by the
distinct-witness condition. Hence the edges correspond one-to-one to the
pairs $(u,(\sigma,\tau))$ of this form.

By equivariance of the neighbor maps, translation by $g\in G$ maps the
edge of $(u,(\sigma,\tau))$ to the edge of $(gu,(\sigma,\tau))$. Since
$G$ acts simply transitively on the middle vertices of each type, the
edge orbits correspond one-to-one to the pairs consisting of a type $x$
and a gadget edge $(\sigma,\tau)$ of $C_x$. The orbit of such a pair joins
$G\times\{\sigma\}$ to $G\times\{\tau\}$. Hence the number of edge orbits
between these two vertex orbits is the number of types $x$ with
$(C_x)_{\tau,\sigma}=1$. Modulo $2$, this number is
$\sum_{x}(C_x)_{\tau,\sigma}$. Counting only the edges witnessed at type $x$ gives $C_x$.
\end{proof}

In particular, when $h$ is even and the same gadget is used at every
middle type, $[A_0]=hC=0$, which is not surjective. By
\cref{lem:orbit-rank-detection}, $A_0$ is not surjective either, so
the original line product does not have full row rank.

To obtain full rank, we must therefore break the cancellation among the
$h$ identical gadget contributions. We will do this by modifying the
gadget at a single middle type. The next subsection constructs the small
additional gadget that will be used for this modification.

\subsection{The full-rank modification}
\label{sec:full-rank-repair}

% The construction of F in the proof of Lemma 3.6, for t=2.
\begin{figure}[!b]
\centering
\begin{tikzpicture}[x=1cm,y=0.7cm]
  \def\Ly{1/1.5,2/0.9,3/0.3,4/-0.3,5/-0.9,6/-1.5}
  % (a) the gadget
  \begin{scope}
    \foreach \i/\y in \Ly \node[Lnode] (a\i) at (0,\y) {};
    \foreach \j/\y in {1/1.2,2/0,3/-1.2} \node[Rnode] (A\j) at (2.2,\y) {};
    \foreach \i/\j in {1/1,2/1,3/2,4/2,5/3,6/3} \draw[gadget] (a\i) -- (A\j);
    \node[figlabel,above=3pt] at (a1.north) {$S_L$};
    \node[figlabel,above=3pt] at (2.2,1.5) {$S_R$};
    \node[figlabel] at (1.1,-2.25) {(a) the gadget $C$};
  \end{scope}
  % (b) t copies of each right vertex, the complement, a perfect matching
  \begin{scope}[xshift=4.6cm]
    \foreach \i/\y in \Ly \node[Lnode] (b\i) at (0,\y) {};
    \foreach \j/\y in {1/1.5,1b/0.9,2/0.3,2b/-0.3,3/-0.9,3b/-1.5}
      \node[Rnode] (B\j) at (2.2,\y) {};
    \begin{scope}[on background layer]
      \foreach \j in {1,2,3}
        \node[group box,fit=(B\j)(B\j b)] {};
    \end{scope}
    % complement: sigma_1,2 miss tau_2,3, etc.
    \foreach \i/\j in {1/2,1/2b,1/3,1/3b,2/2,2/2b,2/3,2/3b,
                       3/1,3/1b,3/3,3/3b,4/1,4/1b,4/3,4/3b,
                       5/1,5/1b,5/2,5/2b,6/1,6/1b,6/2,6/2b}
      \draw[nonedge] (b\i) -- (B\j);
    \foreach \i/\j in {1/2,6/2b,4/1,5/1b,2/3,3/3b}
      \draw[Fedge] (b\i) -- (B\j);
    \node[figlabel,above=3pt] at (b1.north) {$S_L$};
    \node[figlabel,above=3pt] at (2.2,1.65) {$t$ copies of $S_R$};
    \node[figlabel] at (1.1,-2.25) {(b) complement and a matching};
  \end{scope}
  % (c) merging the copies gives F
  \begin{scope}[xshift=9.2cm]
    \foreach \i/\y in \Ly \node[Lnode] (c\i) at (0,\y) {};
    \foreach \j/\y in {1/1.2,2/0,3/-1.2} \node[Rnode] (C\j) at (2.2,\y) {};
    \foreach \i/\j in {1/1,2/1,3/2,4/2,5/3,6/3} \draw[gadget,black!25] (c\i) -- (C\j);
    \foreach \i/\j in {1/2,6/2,4/1,5/1,2/3,3/3} \draw[Fedge] (c\i) -- (C\j);
    \node[figlabel,above=3pt] at (c1.north) {$S_L$};
    \node[figlabel,above=3pt] at (2.2,1.5) {$S_R$};
    \node[figlabel] at (1.1,-2.25) {(c) the matrix $F$};
  \end{scope}
\end{tikzpicture}
\caption{The proof of \cref{lem:complementary-full-rank-factor} for $t=2$.
(a) A gadget with $|S_L|=6$, $|S_R|=3$, and degrees $(1,2)$.
(b) Each right vertex is replaced by $t$ copies (boxed). The dashed
lines are the complement of the gadget, and the red edges are a perfect
matching in it.
(c) Merging the copies gives $F$ (red), next to the gadget (gray). Every
left vertex has one red edge, and every right vertex has $t$.}
\label{fig:F}
\end{figure}
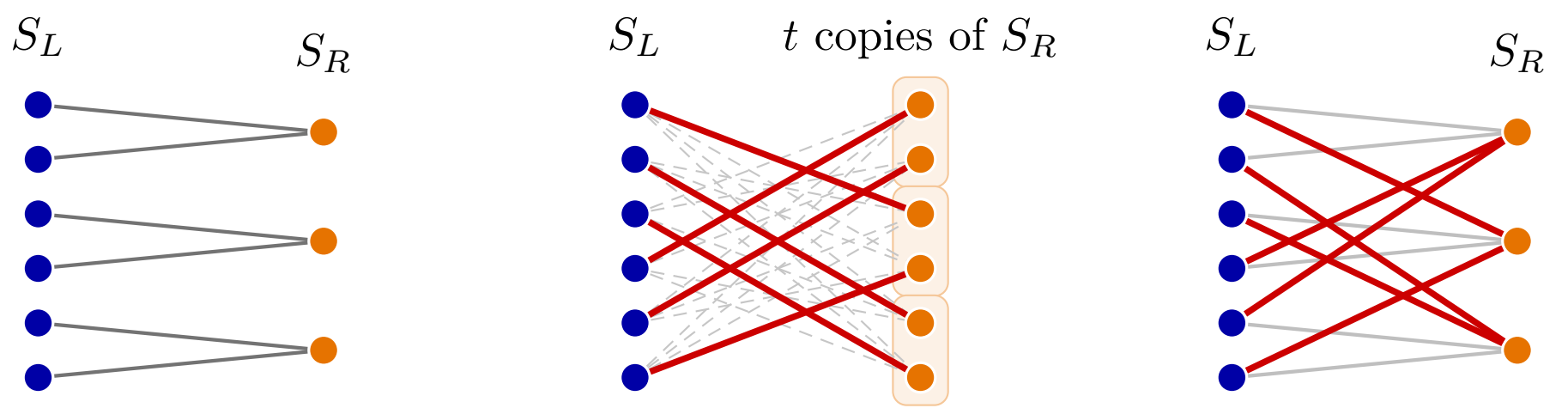

By \cref{lem:line-product-orbit-matrix}, the orbit matrix of $Z_0$
is $hC$, which vanishes since $h$ is even. We break this cancellation by
adding, at the middle vertices of type $0$ only, a matrix
$F\in\mathbb F_2^{S_R\times S_L}$ supported on the nonedges of the gadget.
The orbit matrix then becomes $hC+F=F$, so full row rank of $F$ gives
full row rank of the whole adjacency map, by
\cref{lem:orbit-rank-detection}. The support condition ensures that
no gadget edge is created twice. To keep the graph biregular, with the
smallest possible increase in degree, we ask $F$ to have column weight
one and row weight $t$, where $D_L=tD_R$. Its rows then have nonempty,
pairwise disjoint supports, so $F$ has full row rank automatically. Such
a matrix exists whenever the gadget is not complete.

\begin{lemma}
\label{lem:complementary-full-rank-factor}
Let $C\in\mathbb F_2^{S_R\times S_L}$ be the adjacency matrix of a
simple $(d_L,d_R)$-biregular gadget, where $D_L=tD_R$ for an integer
$t\ge1$ and $d_L<D_R$. Then there is a matrix
$F\in\mathbb F_2^{S_R\times S_L}$, supported on the zero entries of $C$,
with column weight one and row weight $t$. In particular, $F$ has full
row rank.
\end{lemma}

\begin{proof}
Biregularity gives $d_R=td_L$. Replace each vertex of $S_R$ by $t$
copies with the same neighbors, and consider the bipartite complement of
the gadget. It is biregular, with $D_L$ vertices on each side and left
degree $t(D_R-d_L)>0$, hence regular. A
regular bipartite graph of positive degree has a perfect matching.
Merging the copies in such a matching gives $F$
(\cref{fig:F}).
\end{proof}

Let $Z^+$ be the line product with the gadget matrices
\[
 C_0^+=C+F,\qquad C_x^+=C\quad(x\ne0),
\]
and let $A^+\colon\mathbb F_2^L\to\mathbb F_2^R$ be its adjacency map.

\begin{lemma}[Full-rank modification of the line product]
\label{lem:full-rank-line-product-modification}
Assume \cref{ass:rank-free-action} and the distinct-witness
condition~\eqref{eq:distinct-witness}, that $h$ is even, and that the
gadget and $F$ are as in
\cref{lem:complementary-full-rank-factor}. If $Z_0$ is two-sided
$(\alpha,\varepsilon_0)$-lossless, then:
\begin{enumerate}
\item[(i)] $Z^+$ is simple, $G$-invariant, and has degrees
$(hd_L+1,\,t(hd_L+1))$.
\item[(ii)] $A^+$ has full row rank.
\item[(iii)] $Z^+$ is two-sided $(\alpha,\varepsilon_+)$-lossless, where
\[
 \varepsilon_+=\frac{\varepsilon_0hd_L+1}{hd_L+1}
 =\varepsilon_0+\frac{1-\varepsilon_0}{hd_L+1}.
\]
\end{enumerate}
\end{lemma}

\begin{proof}
For (i), note that $0\in H$ and $\phi_\sigma(0)=1$. Hence the edges
added at type $0$ are $(g,\sigma)\sim(g,\tau)$ for $g\in G$ and
$F_{\tau,\sigma}=1$. They form a $G$-invariant set. Since $F$ has column
weight one and row weight $t$, every left vertex gains one edge and every
right vertex gains $t$. With $d_R=td_L$, this gives the stated degrees.
An added edge is not a gadget edge at its own witness, since $F$ is
supported on the nonedges of $C$. By the distinct-witness condition, it
is not witnessed anywhere else either. So $Z^+$ is simple.

For (ii), \cref{lem:line-product-orbit-matrix} gives
$[A^+]=hC+F=F$. This has full row rank, so $A^+$ is surjective by
\cref{lem:orbit-rank-detection}.

For (iii), $Z^+$ is obtained from $Z_0$ by adding edges, so
neighborhoods only grow. For $S\subseteq L$ with $|S|\le\alpha|L|$,
\[
 |N_{Z^+}(S)|\ge|N_{Z_0}(S)|\ge(1-\varepsilon_0)hd_L|S|
 =(1-\varepsilon_+)(hd_L+1)|S|.
\]
The same computation from $R$, with the degrees $hd_R$ and
$hd_R+t=t(hd_L+1)$, gives the bound for $T\subseteq R$.
\end{proof}

Thus it remains to construct lossless line products with distinct
witnesses whose group satisfies \cref{ass:rank-free-action}.
We turn to this in \cref{sec:2group-cubical}.

\section{LPS Ramanujan graphs on 2-groups}
\label{sec:2group-cubical}

The full-rank modification of \cref{sec:full-rank-modification} applies
to the HLMRZ line product once its underlying group is a finite $2$-group
acting freely with a bounded number of orbits
(\cref{ass:rank-free-action}). In HLMRZ, the underlying group
is $\operatorname{PSL}_2(\mathbb F_q)$. This
group comes from the directional graphs of their cubical complexes, which are the
Ramanujan graphs of Lubotzky, Phillips, and Sarnak (LPS)~\cite{LPS88}.
The rest of their construction uses these graphs only through the
Ramanujan bound and through the arithmetic of the integral quaternions
that generate them: unique factorization, and injectivity of reduction on
quaternions of small norm. Neither depends on the vertex group being
$\operatorname{PSL}_2(\mathbb F_q)$. The main task is therefore to
construct analogues of the LPS graphs whose vertex groups are $2$-groups.
As explained in \cref{sec:intro-techniques}, the LPS generators lie in the
congruence subgroup of level $2$, so reducing them modulo $2^m$, instead
of modulo an odd prime $q$, produces $2$-groups.

The section has two parts. \Cref{subsec:2group-lps} constructs the
graphs: for every prime $p\equiv1\pmod4$, the LPS generators of norm $p$,
reduced modulo $2^m$, form a Ramanujan Cayley graph on a $2$-group
(\cref{thm:2group-lps}). For $p\equiv1\pmod8$ the graph is non-bipartite,
and its vertex group is a group $G_m$ of order $2^{3m-4}$, the same for
all such $p$. For $p\equiv5\pmod8$ it is bipartite, on a group
$\widehat G_m$ containing $G_m$ with index two.
\Cref{subsec:2group-complexes} combines several
primes into a cubical complex on $G_m$, as HLMRZ do with the original LPS
graphs (\cref{thm:2group-cubical}). It also shows that two such
complexes with disjoint prime lists satisfy the distinct-witness
condition of the line product (\cref{lem:2group-witnesses}).
\Cref{subsec:2group-line-product-input}
then feeds these complexes into the HLMRZ construction. The arithmetic
inputs, namely the action of the LPS group on the tree and the Ramanujan
property of congruence quotients, are used as black boxes, and we
collect them in Appendix~\ref{app:lps-background}.

\subsection{The graphs}
\label{subsec:2group-lps}

\paragraph{The LPS generators.}
For a commutative ring $R$, write
\[
 \Lips(R)=R+R\qi+R\qj+R\qk,\qquad \qi^2=\qj^2=\qk^2=\qi\qj\qk=-1,
\]
and $\Lips=\Lips(\mathbb Z)$. For $a=a_0+a_1\qi+a_2\qj+a_3\qk$, the
conjugate and norm are $\bar a=a_0-a_1\qi-a_2\qj-a_3\qk$ and
$N(a)=a\bar a=a_0^2+a_1^2+a_2^2+a_3^2$. Congruences are coefficientwise.
For a positive integer $n\equiv1\pmod4$, put
\[
 A(n)=\{a\in\Lips:N(a)=n,\ a_0\text{ odd}\}/\{\pm1\}.
\]
Since squares are $0$ or $1$ modulo $4$, a quaternion of norm $n$ has
exactly one odd coordinate. Hence every $a\in A(n)$ has even pure
coordinates, that is, $a\equiv1\pmod2$. For a prime $p\equiv1\pmod4$,
the set $A(p)$ has $p+1$ elements, and their representatives with
$a_0>0$ are the LPS generators (\cref{lem:lps-integral-products}). LPS
reduce them modulo an odd prime $q$, whereas we reduce them modulo $2^m$.

\paragraph{The groups $G_m$.}
Fix $m\ge3$ and put $R_m=\mathbb Z/2^m\mathbb Z$. The set
$U_m=1+2\Lips(R_m)$ of quaternions with odd real part and even pure part
is a group: the two-sided ideal $2\Lips(R_m)$ is nilpotent, so $1+2y$ is
inverted by a finite geometric series. The scalar units $R_m^\times$ form
a central subgroup.

\begin{definition}[The groups $G_m$ and $\widehat G_m$]
\label{def:2group-groups}
\[
 \widehat G_m=U_m/R_m^\times,\qquad
 G_m=\{[x]\in\widehat G_m:N(x)\equiv1\pmod8\}.
\]
\end{definition}

The norm condition is well defined because $c^2\equiv1\pmod8$ for odd
$c$, and it defines a subgroup because the norm is multiplicative.
Dividing by the real part, every class in $\widehat G_m$ has a unique
representative
\[
 1+2u,\qquad u=u_1\qi+u_2\qj+u_3\qk,\qquad
 u_t\in\mathbb Z/2^{m-1}\mathbb Z,
\]
so $|\widehat G_m|=2^{3(m-1)}$. We call these representatives
\emph{normal forms}. Every integer satisfies $x^2\equiv x\pmod2$,
hence $4x^2\equiv4x\pmod8$, and therefore
$N(1+2u)\equiv1+4(u_1+u_2+u_3)\pmod8$. Thus $[1+2u]\in G_m$ exactly when
$u_1+u_2+u_3$ is even, which holds for half of the normal forms:
\begin{equation}
\label{eq:2group-size}
 [\widehat G_m:G_m]=2,\qquad |G_m|=2^{3m-4}.
\end{equation}
In normal forms, the group law is
\begin{equation}
\label{eq:2group-normal-product}
 (1+2u)(1+2v)=(1-4\,u\cdot v)+2(u+v+2\,u\times v),
\end{equation}
where $u\cdot v$ and $u\times v$ are the usual dot and cross products of
the coefficient vectors $(u_1,u_2,u_3)$ and $(v_1,v_2,v_3)$. This follows
from the identity $uv=-u\cdot v+u\times v$ for pure quaternions, and
dividing by the odd real part gives the normal form of the product.
Inverses are $[1+2u]^{-1}=[1-2u]$.

\paragraph{Reduction modulo $2^m$.}
Let $\mathbb Z_2$ be the ring of $2$-adic integers. Every
$x\in1+2\Lips(\mathbb Z_2)$ is invertible, with inverse $N(x)^{-1}\bar x$,
since $N(x)\equiv1\pmod4$ is a unit. Reduction modulo $2^m$ followed by
the quotient by scalars is a surjective homomorphism
\[
 \pi_m\colon 1+2\Lips(\mathbb Z_2)\longrightarrow\widehat G_m.
\]
It kills the scalars $\mathbb Z_2^\times$. Hence
$\pi_m(\bar a)=\pi_m(a)^{-1}$, as $a\bar a=N(a)$ is a scalar, and
$\pi_m(a)\in G_m$ if and only if $N(a)\equiv1\pmod8$. For an odd prime $p$
we have $\mathbb Z[1/p]\subseteq\mathbb Z_2$, so $\pi_m$ applies to every
quaternion over $\mathbb Z[1/p]$ congruent to $1$ modulo $2$, and in
particular to $A(n)$. Moreover, $\pi_m$ is continuous for the $2$-adic
topology on $\Lips(\mathbb Z_2)\cong\mathbb Z_2^4$ and the discrete
topology on $\widehat G_m$, since it is locally constant: $\pi_m(x)$
depends only on $x$ modulo $2^m$, so $\pi_m$ is constant on the open
neighborhood $x+2^m\Lips(\mathbb Z_2)$ of each $x$. For a prime
$p\equiv1\pmod4$, put
\[
 A_p^{(m)}=\pi_m(A(p)).
\]
If $p\equiv1\pmod8$, then $A_p^{(m)}\subseteq G_m$.

\paragraph{The theorem.}
Recall that a connected $k$-regular graph is \emph{Ramanujan} if every adjacency
eigenvalue $\lambda\ne\pm k$ satisfies $|\lambda|\le2\sqrt{k-1}$. We prove the
following theorem:

\begin{theorem}[LPS Ramanujan graphs on $2$-groups]
\label{thm:2group-lps}
Let $p\equiv1\pmod4$ be prime and let $m\ge3$ satisfy $2^m>2\sqrt p$.
Then $A_p^{(m)}$ consists of $p+1$ distinct nonidentity elements, and
$\operatorname{Cay}(\langle A_p^{(m)}\rangle;A_p^{(m)})$ is a simple,
connected, $(p+1)$-regular Ramanujan graph. Moreover:
\begin{enumerate}
\item[(i)] If $p\equiv1\pmod8$, then $\langle A_p^{(m)}\rangle=G_m$, the
graph is non-bipartite, and its girth is at least $\frac23\log_p|G_m|$.
\item[(ii)] If $p\equiv5\pmod8$, then
$\langle A_p^{(m)}\rangle=\widehat G_m$, the graph is bipartite with parts
$G_m$ and $\widehat G_m\setminus G_m$, and its girth is at least
$\frac43\log_p|\widehat G_m|$.
\end{enumerate}
\end{theorem}

This is the analogue of the LPS theorem~\cite{LPS88}: the pair
$G_m\subseteq\widehat G_m$ plays the role of
$\operatorname{PSL}_2(\mathbb F_q)\subseteq\operatorname{PGL}_2(\mathbb F_q)$,
and the condition $p\equiv1\pmod8$ replaces the condition that $p$ is a
square modulo $q$. Indeed, for $m\ge3$ an odd integer is a square modulo
$2^m$ exactly when it is $1$ modulo $8$. Odd squares are $1$ modulo $8$,
and conversely every $2$-adic unit that is $1$ modulo $8$ is a $2$-adic
square~\cite[Chapter~II, \S3.3, Theorem~4]{Ser73}.

\paragraph{Strategy of the proof.}
Two ingredients of the LPS construction carry over unchanged to the
level $2^m$, and we use them as black boxes. First, the LPS group is free
and acts simply transitively on the $(p+1)$-regular tree, so every
congruence quotient of the tree is a Cayley graph
(\cref{thm:lps-tree}). Second, every congruence quotient of level prime
to $p$ is Ramanujan (\cref{thm:lps-congruence-ramanujan}). This is the
deep input, resting on Deligne's proof of the Ramanujan--Petersson
conjecture. What requires a new argument is the identification of the
vertex group. At an odd prime $q$ the quaternion algebra splits, so
reduction modulo $q$ takes values in a matrix group over $\mathbb F_q$,
and LPS identify the image with $\operatorname{PSL}_2(\mathbb F_q)$ or
$\operatorname{PGL}_2(\mathbb F_q)$. At $2$ the algebra is ramified, and
no such matrix description is available. We use strong approximation, as
in the LPS setting, to pass to a $2$-adic group, and then compute its
image modulo $2^m$ directly (\cref{prop:2group-common-image}). Finally,
an elementary size argument shows that the $p+1$ generators remain
distinct and nontrivial after reduction (\cref{lem:2group-nondegeneracy}).

The following lemma is stated for general $n$, since it is also applied
to products of several primes in \cref{subsec:2group-complexes}.

\begin{lemma}[Non-degeneracy]
\label{lem:2group-nondegeneracy}
Let $n\equiv1\pmod4$ be a positive integer that is not a square, and let
$m\ge3$ satisfy $2^m>2\sqrt n$. Then $\pi_m$ is injective on $A(n)$, and
$1\notin\pi_m(A(n))$.
\end{lemma}

\begin{proof}
Let $a,b\in\Lips$ represent elements of $A(n)$. Their real parts are
odd, hence nonzero, so their pure coordinates are smaller than
$\sqrt n<2^{m-1}$ in absolute value. Suppose that $\pi_m(a)=\pi_m(b)$, so
$a\equiv cb\pmod{2^m}$ for an odd integer $c$. Taking norms gives
$c^2\equiv1\pmod{2^m}$, since $n$ is odd. One of the consecutive even
numbers $c-1,c+1$ is $2$ modulo $4$, so the other is divisible by
$2^{m-1}$, that is, $c\equiv\pm1\pmod{2^{m-1}}$. Replacing $b$ by $-b$, we may assume
$c\equiv1\pmod{2^{m-1}}$, and write $c=1+2^{m-1}k$. Since the pure
coordinates $b_t$ are even, $cb_t\equiv b_t\pmod{2^m}$, so
$a_t\equiv cb_t\equiv b_t\pmod{2^m}$ for $t=1,2,3$. As
$|a_t|,|b_t|<2^{m-1}$, this gives $a_t=b_t$. Since $N(a)=N(b)$, it
follows that $a_0=\pm b_0$. For the real coordinates, $a\equiv cb$ gives
$a_0\equiv cb_0\pmod{2^m}$, and $c\equiv1\pmod{2^{m-1}}$ then gives
$a_0\equiv b_0\pmod{2^{m-1}}$. If $a_0=-b_0$, this says
$2b_0\equiv0\pmod{2^{m-1}}$, so $b_0$ is even since $m\ge3$, a
contradiction. Hence $a=b$.

If $\pi_m(a)=1$, the pure coordinates of $a$ are divisible by $2^m$, so
they vanish, and $n=a_0^2$, a contradiction.
\end{proof}

\paragraph{The tree quotient.}
Fix a prime $p\equiv1\pmod4$. Let $\Lambda_p$ be the subgroup of the
projective group $\Lips(\mathbb Z[1/p])^\times/\mathbb Z[1/p]^\times$
generated by the classes of $A(p)$, where
$\mathbb Z[1/p]^\times=\{\pm p^z:z\in\mathbb Z\}$. By \cref{thm:lps-tree}, $\Lambda_p$
is free on $(p+1)/2$ generators, with $A(p)$ as symmetric generating set,
and
\[
 T_{p+1}\cong\operatorname{Cay}(\Lambda_p;A(p))
\]
is the $(p+1)$-regular tree. Let $\Lambda_p^+$ be the index-two subgroup
of words of even length. The representatives of classes in $\Lambda_p$
are congruent to $1$ modulo $2$ (\cref{thm:lps-tree}(i)) and differ by
scalars, so $\pi_m$ induces a homomorphism
$\pi_m\colon\Lambda_p\to\widehat G_m$. Its image is
$\langle A_p^{(m)}\rangle$, and its kernel is the congruence subgroup
$\Gamma_p(2^m)$ of~\eqref{eq:lps-congruence-subgroup}. Dividing the tree
by this normal subgroup gives, as in~\eqref{eq:lps-congruence-cayley},
\begin{equation}
\label{eq:2group-tree-quotient}
 X^{p,2^m}:=\Gamma_p(2^m)\backslash T_{p+1}
 \cong\operatorname{Cay}\bigl(\pi_m(\Lambda_p);A_p^{(m)}\bigr).
\end{equation}
This graph is Ramanujan by \cref{thm:lps-congruence-ramanujan}. It
remains to identify the group $\pi_m(\Lambda_p)$.

\paragraph{The vertex group.}
For a commutative ring $R$, let $\Lips^1(R)$ be the group of norm-one
elements of $\Lips(R)$, and let $\Lips^1(R;2)$ be its subgroup of
elements congruent to $1$ modulo $2$.

\begin{proposition}[Common image]
\label{prop:2group-common-image}
For every prime $p\equiv1\pmod4$ and every $m\ge3$,
\[
 \pi_m(\Lambda_p^+)=G_m.
\]
\end{proposition}

\begin{proof}
We prove the three equalities in
\begin{equation}
\label{eq:2group-dense-image}
 \pi_m(\Lambda_p^+)
 =\pi_m\bigl(\Lips^1(\mathbb Z[1/p];2)\bigr)
 =\pi_m\bigl(\Lips^1(\mathbb Z_2;2)\bigr)
 =G_m.
\end{equation}
The first passes from $\Lambda_p^+$ to norm-one quaternions, the second
uses strong approximation to pass to the $2$-adic group, and the third
computes the image of that group directly.

\emph{The first equality.}
A word $w$ of length $2k$ in integral representatives of $A(p)$ has norm
$p^{2k}$, so $p^{-k}w\in\Lips^1(\mathbb Z[1/p];2)$ represents the same
class. Conversely, let $x\in\Lips^1(\mathbb Z[1/p];2)$. By
\cref{thm:lps-tree}(i), $[x]\in\Lambda_p$, so $x=\pm p^z a_1\cdots a_k$
with integral representatives $a_i$ of elements of $A(p)$. Taking norms
gives $1=p^{2z+k}$, so $k$ is even and $[x]\in\Lambda_p^+$. Here the
parity of the word length is well defined, since $\Lambda_p$ is free on
$A(p)$ up to inverses (\cref{thm:lps-tree}(ii)). Since $\pi_m$
kills scalars, the two images agree.

\emph{The second equality.}
The inclusion $\subseteq$ is clear. For the reverse inclusion, we use
the strong approximation theorem~\cite{Kne66}, \cite[Theorem~7.12]{PR94},
in the form given in \cite{LS03}. If $\mathbf G$ is a connected,
simply connected, $\mathbb Q$-simple algebraic group over $\mathbb Q$
and $\mathbf G(\mathbb Z[1/p])$ is infinite, then $\mathbf G(\mathbb Z[1/p])$
is dense in $\prod_{\ell\ne p}\mathbf G(\mathbb Z_\ell)$. We apply this to
the norm-one group $\mathbf G=\operatorname{SL}_1(D)$ of the Hamilton
quaternion algebra $D$, with the integral structure given by the Lipschitz
order, so that $\mathbf G(\mathbb Z[1/p])=\Lips^1(\mathbb Z[1/p])$ and
$\mathbf G(\mathbb Z_2)=\Lips^1(\mathbb Z_2)$. Over an algebraic closure,
$\mathbf G$
becomes $\operatorname{SL}_2$, so it is connected, simply connected, and
$\mathbb Q$-simple. The group $\Lips^1(\mathbb Z[1/p])$ is infinite,
since by the proof of the first equality its subgroup $\Lips^1(\mathbb Z[1/p];2)$
maps onto the infinite group $\Lambda_p^+$. Projecting to the factor
$\ell=2$, we find that $\Lips^1(\mathbb Z[1/p])$ is dense in
$\Lips^1(\mathbb Z_2)$. As $\Lips^1(\mathbb Z_2;2)$ is an open
subgroup, $\Lips^1(\mathbb Z[1/p];2)$ is dense in it. Since
$\pi_m$ is continuous and $\widehat G_m$ is finite, every nonempty fiber
of $\pi_m$ on $\Lips^1(\mathbb Z_2;2)$ is open, and hence meets this
dense subgroup.

\emph{The third equality.}
Norm-one elements map into $G_m$. Conversely, let $[x]\in G_m$, and lift
$x$ to $\widetilde x\in1+2\Lips(\mathbb Z_2)$. Then
$N(\widetilde x)\equiv1\pmod8$, so $N(\widetilde x)=c^2$ for some
$c\in\mathbb Z_2^\times$~\cite[Chapter~II, \S3.3, Theorem~4]{Ser73}.
Since $c$ is
odd, $\widetilde x/c\in\Lips^1(\mathbb Z_2;2)$, and
$\pi_m(\widetilde x/c)=\pi_m(\widetilde x)=[x]$.
\end{proof}

We are now in a position to prove \cref{thm:2group-lps}, by combining
the above claims with
the Ramanujan property of congruence quotients, in the form of
\cref{thm:lps-congruence-ramanujan}.

\begin{proof}[Proof of \cref{thm:2group-lps}]
By \cref{lem:2group-nondegeneracy} with $n=p$, the set $A_p^{(m)}$
consists of $p+1$ distinct nonidentity elements. It is inverse-closed,
since $\pi_m(\bar a)=\pi_m(a)^{-1}$. Thus the Cayley graph is simple and
$(p+1)$-regular, and by~\eqref{eq:2group-tree-quotient} it is
$X^{p,2^m}$.

By \cref{prop:2group-common-image}, the vertex group $\pi_m(\Lambda_p)$
contains $G_m$, which has index two in $\widehat G_m$. If
$p\equiv1\pmod8$, the generators lie in $G_m$, so $\pi_m(\Lambda_p)=G_m$.
If $p\equiv5\pmod8$, they lie in the other coset, so
$\pi_m(\Lambda_p)=\widehat G_m$ and every edge joins the two cosets.

Finally, \cref{thm:lps-congruence-ramanujan}, applied with $N=2^m$, shows
that $X^{p,2^m}$ is Ramanujan. Its two cases depend on whether $p$ is a
square modulo $2^m$, which holds exactly when $p\equiv1\pmod8$, as noted
after the statement of the theorem. Hence the graph is non-bipartite in
case~(i) and bipartite in case~(ii), and the girth bounds follow from the
same theorem, using~\eqref{eq:2group-size}.
\end{proof}

\subsection{The cubical complexes}
\label{subsec:2group-complexes}

Since the group $G_m$ does not depend on $p$, the graphs of several
primes $p\equiv1\pmod8$ live on the same group, and we can combine them
into a cubical complex exactly as HLMRZ do \cite[Section~4]{HLMRZ25}.
The mechanism is the following. Reduction modulo $2^m$ is injective on
the integral quaternions involved, which have bounded norm. Therefore the
multiplicative structure that unique factorization gives the integral
generators survives in $G_m$: the generating sets of different primes
commute, and their ordered products are all distinct. Consequently, the
operator of a mixed direction factors into one-direction operators, and
the Ramanujan bounds multiply. Compared with HLMRZ, the same integral
generators are reduced modulo $2^m$ instead of an odd prime $q$, and the
separation condition~\eqref{eq:2group-separation} below replaces their
condition $q>2\sqrt{\prod_ip_i}$. Their quadratic-residue condition
becomes $p\equiv1\pmod8$. Once the primes are fixed, increasing $m$ gives an infinite
family.

Let $A_1,\ldots,A_\ell$ be cubical generating sets of a finite group $G$
(\cref{sec:tripartite-line-product}). For $\varnothing\ne I\subseteq[\ell]$,
write $A_I=\prod_{i\in I}A_i$ and $d_I=\prod_{i\in I}|A_i|$, and let
$\mathbf1_I\in\mathbb F_2^\ell$ be the indicator vector of $I$. Since the
$A_i$ commute, $A_I$ does not depend on the order of the factors. The
bipartite graph between types $y$ and $y+\mathbf1_I$ has edges
\[
 (g,y)\sim(ga,\,y+\mathbf1_I),\qquad a\in A_I.
\]
The complex is \emph{$\alpha$-expanding} if, for every nonempty $I$ and
every $y$, the second largest eigenvalue of the adjacency matrix of this
graph is at most $\alpha\sqrt{d_I}$.
Let $(T_If)(g)=\sum_{a\in A_I}f(ga)$ on $\ell^2(G)$. The adjacency
eigenvalues of this bipartite graph are $\pm$ the singular values of
$T_I$. So the complex is $\alpha$-expanding exactly when, for every
nonempty $I$, the operator $T_I$ has norm at most $\alpha\sqrt{d_I}$ on
the orthogonal complement $\mathbf1^\perp$ of the constant functions.

From now on, let $\ell\ge1$, let $p_1,\ldots,p_\ell$ be distinct primes
with $p_i\equiv1\pmod8$, and let $m\ge3$ satisfy
\begin{equation}
\label{eq:2group-separation}
 2^m>2\sqrt{p_1p_2\cdots p_\ell}.
\end{equation}
The next lemma transfers the unique factorization of integral quaternions
to $G_m$. For $I\subseteq[\ell]$, put
\[
 n_I=\prod_{i\in I}p_i,\qquad
 A_I^{(m)}=\prod_{i\in I}A_{p_i}^{(m)}.
\]

\begin{lemma}
\label{lem:2group-mixed-products}
For every nonempty $I\subseteq[\ell]$ and every ordering of $I$,
multiplication followed by reduction is a bijection
\[
 \prod_{i\in I}A(p_i)\xrightarrow{\ \sim\ }A_I^{(m)}=\pi_m\bigl(A(n_I)\bigr).
\]
In particular, $A_I^{(m)}$ does not depend on the ordering, and
$|A_I^{(m)}|=\prod_{i\in I}(p_i+1)$.
\end{lemma}

\begin{proof}
For every ordering, multiplication is a bijection from
$\prod_{i\in I}A(p_i)$ onto $A(n_I)$, by unique factorization of integral
quaternions (\cref{lem:lps-integral-products}). The
integer $n_I>1$ is squarefree, hence not a square, and
\eqref{eq:2group-separation} gives $2^m>2\sqrt{n_I}$. By
\cref{lem:2group-nondegeneracy}, $\pi_m$ is injective on $A(n_I)$.
\end{proof}

We prove the following theorem.

\begin{theorem}[Ramanujan cubical complexes on $2$-groups]
\label{thm:2group-cubical}
Assume~\eqref{eq:2group-separation}. Then the sets
$A_{p_1}^{(m)},\ldots,A_{p_\ell}^{(m)}$ are cubical generating sets of
$G_m$, with $|A_{p_i}^{(m)}|=p_i+1$, and the decorated cubical complex
\[
 Y_m=\operatorname{Cay}(G_m;A_{p_1}^{(m)},\ldots,A_{p_\ell}^{(m)})
\]
is $2^\ell$-expanding.
\end{theorem}

\begin{proof}
We follow the argument of HLMRZ. By~\eqref{eq:2group-separation},
$2^m>2\sqrt{p_i}$ for every $i$. So by
\cref{thm:2group-lps}, each $A_{p_i}^{(m)}$ is inverse-closed, lies in
$G_m$, and has $p_i+1$ elements. By \cref{lem:2group-mixed-products},
\[
 A_{p_i}^{(m)}A_{p_j}^{(m)}=\pi_m\bigl(A(p_ip_j)\bigr)
 =A_{p_j}^{(m)}A_{p_i}^{(m)}\qquad(i\ne j),
\]
and $|A_{p_1}^{(m)}\cdots A_{p_\ell}^{(m)}|=\prod_i(p_i+1)
=\prod_i|A_{p_i}^{(m)}|$. Thus these are cubical generating sets.

For expansion, fix one of the primes $p=p_i$, and let $T_{p,m}$ be the
adjacency operator of $\operatorname{Cay}(G_m;A_p^{(m)})$. It is
self-adjoint, since $A_p^{(m)}$ is inverse-closed. The graph is
connected, non-bipartite, and Ramanujan (\cref{thm:2group-lps}(i)). Hence
$\pm(p+1)$ are not eigenvalues of $T_{p,m}$ on $\mathbf1^\perp$, and all
its eigenvalues there have absolute value at most $2\sqrt p$. Now let
$I=\{i_1<\cdots<i_k\}$ be nonempty. By \cref{lem:2group-mixed-products},
every element of $A_I^{(m)}$ is a product $a_1\cdots a_k$ with
$a_j\in A_{p_{i_j}}^{(m)}$ in exactly one way. Hence
$T_I=T_{p_{i_1},m}\cdots T_{p_{i_k},m}$. Every factor preserves
$\mathbf1^\perp$, so
\[
 \bigl\|T_I|_{\mathbf1^\perp}\bigr\|
 \le\prod_{i\in I}2\sqrt{p_i}
 \le2^{|I|}\sqrt{d_I}
 \le2^\ell\sqrt{d_I}.
 \qedhere
\]
\end{proof}

The tripartite line product uses two such complexes on the same group
(\cref{sec:tripartite-line-product}). Let $P^L$ and $P^R$ be disjoint
lists of $h$ primes congruent to $1$ modulo $8$, and let $m$
satisfy~\eqref{eq:2group-separation} for the union of the $2h$ primes.
\Cref{thm:2group-cubical} gives two $2^h$-expanding complexes $Y_L,Y_R$
on $G_m$, with the common vertex set $G_m\times\mathbb F_2^h$. A
superscript $\bullet\in\{L,R\}$ distinguishes their products of
generating sets $A_I^{(m),\bullet}$.

\begin{lemma}
\label{lem:2group-witnesses}
In this setting, a cube of $Y_L$ and a cube of $Y_R$ share at most one
vertex. In particular, the distinct-witness
condition~\eqref{eq:distinct-witness} holds for all signature sets.
\end{lemma}

\begin{proof}
Suppose that a cube $f$ of $Y_L$ and a cube $f'$ of $Y_R$ share two
vertices. A cube has exactly one vertex of each type, so these are
$(f_x,x)=(f'_x,x)$ and $(f_y,y)=(f'_y,y)$ for two distinct types $x,y$.
The element $f_x^{-1}f_y$ is the product of the edge labels of $f$ along
a path from $x$ to $y$ that changes each coordinate in
$I=\operatorname{supp}(x+y)$ once, so it lies in $A_I^{(m),L}$. In the
same way, $f_x^{\prime-1}f'_y=f_x^{-1}f_y$ lies in $A_I^{(m),R}$. Hence
$A_I^{(m),L}\cap A_I^{(m),R}$ is nonempty. By
\cref{lem:2group-mixed-products}, there are $a\in A(n)$ and
$b\in A(n')$ with $\pi_m(a)=\pi_m(b)$, where $n,n'>1$ are coprime
squarefree products of left and right primes, respectively. Then
$\pi_m(a\bar b)=\pi_m(a)\pi_m(b)^{-1}=1$, and $a\bar b$ represents an
element of $A(nn')$. But $nn'$ is not a square, and $2^m>2\sqrt{nn'}$ by
the separation condition for the union. This contradicts
\cref{lem:2group-nondegeneracy}.
\end{proof}

\section{Full-rank lossless graphs and the quantum code construction}
\label{codingcompletion:subsection}

We now complete the construction. First we feed the cubical complexes of
\cref{sec:2group-cubical} into the HLMRZ construction
(\cref{thm:2group-lossless}). The modification of
\cref{lem:full-rank-line-product-modification} then gives
full-rank lossless graphs with a free action of $G_m$, which proves
\cref{thm:intro-full-rank-lossless}. Finally, we plug two such graphs
into the Lin--Hsieh balanced product and prove
\cref{thm:intro-independent-checks}. All ranks in this
section are over $\mathbb F_2$.

\subsection{Lossless graphs with a free 2-group action}
\label{subsec:2group-line-product-input}

HLMRZ prove their main theorem (\cref{thm:HLMRZ}) from two
ingredients. The first is a pair
of structured base graphs, namely incidence graphs $I_L,I_R$ as in
\cref{sec:tripartite-line-product} with three combinatorial properties
(\cite[Lemma~2.8]{HLMRZ25}). The second is a random gadget with local
expansion and spreading properties (\cite[Lemma~2.9]{HLMRZ25}). The
argument of \cite[Section~2.2]{HLMRZ25} uses the base graphs only through
these three properties. Their proof in
\cite[Sections~3.1 and~3.2]{HLMRZ25} uses the two cubical complexes only
through the conclusions of the HLMRZ theorem on Ramanujan cubical
complexes over $\operatorname{PSL}_2(\mathbb F_q)$
(\cite[Theorem~3.5]{HLMRZ25}). Namely, the two complexes live on the
same group, their generating sets are cubical of sizes $p_i+1$ for primes
$p_i$ in an interval $[x,2x]$, and both complexes are $2^h$-expanding. \Cref{thm:2group-cubical} gives the same conclusions on $G_m$, for
primes congruent to $1$ modulo $8$. The construction therefore applies
verbatim, with $\operatorname{PSL}_2(\mathbb F_q)$ replaced by $G_m$ and
the condition $q>2\sqrt{\prod_ip_i}$ replaced
by~\eqref{eq:2group-separation}.

\begin{theorem}[Lossless line products with a free $2$-group action]
\label{thm:2group-lossless}
Let $t\ge1$ be an integer and let $\varepsilon\in(0,1)$. There is $d_0$
such that for every integer $d\ge d_0$ there are a power of two $h\ge2$, two
disjoint lists $P^L,P^R$ of $h$ primes congruent to $1$ modulo $8$, an
integer $D_R>d$, a constant $c>0$, and an integer $m_0$ with the
following property. For $m\ge m_0$, let $Y_L,Y_R$ be the complexes of
\cref{thm:2group-cubical} on $G_m$ for $P^L$ and $P^R$. There are
signature sets $S_L,S_R$ with $|S_L|=tD_R$ and $|S_R|=D_R$, and a gadget
$C$ of degrees $(d,td)$, such that
$Z_0=Z(I(Y_L,S_L),I(Y_R,S_R);C)$ satisfies the following:
\begin{enumerate}
\item[(i)] $Z_0$ is simple, $(hd,htd)$-biregular, and two-sided
$(c,\varepsilon)$-lossless.
\item[(ii)] $Z_0$ satisfies the distinct-witness
condition~\eqref{eq:distinct-witness}.
\item[(iii)] Left translation is a free action of the $2$-group $G_m$,
with $|S_L|$, $|S_R|$, and $h$ orbits on $L$, $R$, and $M$.
\item[(iv)] The sets $S_L,S_R$, the gadget $C$, and the neighbors of a
vertex of $Z_0$ are computable in time polynomial in $m$.
\end{enumerate}
\end{theorem}

\begin{proof}
For (i), we run the proof of the HLMRZ main theorem
(\cite[Theorem~2.2]{HLMRZ25}, stated as \cref{thm:HLMRZ}) verbatim on
the complexes of \cref{thm:2group-cubical}, with ratio $\beta=1/t$ and
degrees $(d,td)$. Primes congruent to $1$ modulo $8$ in $[x,2x]$ exist
for every large $x$~\cite{BMOR18}, and $m_0$ is the least $m\ge3$
satisfying~\eqref{eq:2group-separation} for the union of $P^L$ and $P^R$.
The proof takes $D_L=tD_R$, and its degrees satisfy $td\le D^{3/8}$
with $D=(t+1)D_R$, so $D_R>d$ for large $d$. The density $c$ depends
only on the degrees and the loss. Simplicity and (ii) follow from
\cref{lem:2group-witnesses}, and (iii) holds since $G_m$ is a $2$-group
acting freely by left translation, with orbits indexed by $S_L$, $S_R$,
and $H$.

For (iv), the signatures are the square-consistent labelings of the
cube by the generating sets. Finding them, the normalized cubes
$\phi_\sigma$, and the choices of $S_L$, $S_R$, and $C$ is a finite
search whose size depends only on the primes and the degrees. Each step
is a fixed number of operations in $G_m$, each polynomial in $m$
by~\eqref{eq:2group-normal-product}. The neighbors of $(g,\sigma)$ are
$\bigl(g\phi^L_\sigma(x)\phi^R_\tau(x)^{-1},\tau\bigr)$ for $x\in H$ and
gadget edges $(\sigma,\tau)$, and the right neighbors are similar.
\end{proof}

\subsection{Full-rank lossless expanders}
\label{codingcompletion:graphs}

\begin{theorem}[Full-rank lossless expanders, precise form of \cref{thm:intro-full-rank-lossless}]
\label{codingcompletion:prop:graphs}
For every integer $t\ge1$, every $\varepsilon\in(0,1)$, and every positive
integer $a_{\min}$, there are positive integers $u,a$, a constant $c>0$,
and an integer $m_0$, with $a\ge a_{\min}$, such that for every $m\ge m_0$
there is a simple bipartite graph
\[
 Z_m=(G_m\times[tu],\,G_m\times[u],\,E_m)
\]
with degrees $(a,ta)$, two-sided $(c,\varepsilon)$-losslessness, and a
full-row-rank binary adjacency map
\[
 \mathbb F_2^{G_m\times[tu]}\longrightarrow
 \mathbb F_2^{G_m\times[u]}.
\]
Left translation is a free $G_m$-action. Neighbors on either side can be
enumerated in time polynomial in $m=O(\log|G_m|)$.
\end{theorem}

\begin{proof}
Apply \cref{thm:2group-lossless} with $t$, loss $\varepsilon/2$, and an
integer $d$ with $d\ge d_0$, $d\ge a_{\min}$, and $d\ge2/\varepsilon$.
It gives a power of two $h\ge2$, an integer $D_R>d$, a constant $c>0$,
and an integer $m_0$. Put $u=D_R$ and $a=hd+1$.

Let $m\ge m_0$, and let $S_L,S_R$ and $C$ be as in
\cref{thm:2group-lossless}. Identify $S_L$ with $[tu]$ and $S_R$ with
$[u]$. By \cref{thm:2group-lossless}, the line product $Z_0$ is
simple and two-sided $(c,\varepsilon/2)$-lossless, satisfies the
distinct-witness condition~\eqref{eq:distinct-witness}, and satisfies
\cref{ass:rank-free-action}. Since $h$ is even and the gadget
is simple with $|S_L|=t|S_R|$ and $d<|S_R|$, the hypotheses of
\cref{lem:full-rank-line-product-modification} hold. Let $Z_m$ be
the modified graph $Z^+$. By that lemma, $Z_m$ is simple with degrees
$(hd+1,t(hd+1))=(a,ta)$, its adjacency map has full row rank, and left
translation by $G_m$ acts freely. It is two-sided lossless with the same
density $c$ and with loss
\[
 \varepsilon_+
 =\frac\varepsilon2+\frac{1-\varepsilon/2}{a}
 \le\frac\varepsilon2+\frac1a\le\varepsilon,
\]
since $a>d\ge2/\varepsilon$. Moreover $a>d\ge a_{\min}$.

Neighbors in $Z_0$ are computable in time polynomial in $m$ by
\cref{thm:2group-lossless}(iv). By the proof of
\cref{lem:full-rank-line-product-modification}, the added edges are
$(g,\sigma)\sim(g,\tau)$ with $F_{\tau,\sigma}=1$. The matrix $F$ of
\cref{lem:complementary-full-rank-factor} is found from $C$ by a
finite matching computation, whose size depends only on $u$, $t$, and
$d$.
\end{proof}

\Cref{thm:intro-full-rank-lossless} follows. The graphs $Z_m$ have
$(t+1)u|G_m|$ vertices and $(t+1)u$ orbits under the free action of
$G_m$, so $m=O(\log|G_m|)$ is logarithmic in the number of vertices.

\subsection{The quantum codes}
\label{codingcompletion:assembly}

Lin and Hsieh construct good quantum LDPC codes from two-sided lossless
expanders with a free group action, assuming such expanders exist
(\cite[Conjecture~10 and Theorem~9]{LH22Quantum}).
\Cref{codingcompletion:prop:graphs} provides them, with full rank in
addition. We prove \cref{thm:intro-independent-checks} by running their
construction with these graphs. Full rank of the input graphs gives
independent checks by \cref{thm:LH-rank-inheritance}, and the small-set
estimate of Lin and Hsieh gives small-set soundness.

\paragraph{The small-set estimate.}
Let $\Gamma_A=(A_0,A_1,E_A)$ and $\Gamma_B=(B_0,B_1,E_B)$ be two-sided
$(c_1,\varepsilon_1)$- and $(c_2,\varepsilon_2)$-lossless bipartite graphs
with a common free $G$-action. Let $\Gamma_A$ have degrees $D_1$ on $A_0$
and $D'_1$ on $A_1$, and let $\Gamma_B$ have degrees $D_2$ on $B_0$ and
$D'_2$ on $B_1$. In the code of \cref{app:LH}, the qubits are
$V_{10}\sqcup V_{01}$. For
$y=(y_{10},y_{01})\in\mathbb F_2^{V_{10}}\oplus\mathbb F_2^{V_{01}}$, put
\[
 |y|_w=\frac{|y_{10}|}{D_1}+\frac{|y_{01}|}{D_2},
\]
and call $y$ \emph{locally minimal} if adding a row of $H_X$ to $y$ does
not decrease $|y|_w$. Suppose that $\varepsilon_1\le\varepsilon$,
$\varepsilon_1D_1\le D'_1$, and $\varepsilon_2D_1\le\varepsilon$. Then
every locally minimal $y$ with $|y_{10}|,|y_{01}|<\kappa|G|$ satisfies
\begin{equation}
\label{codingcompletion:eq:LH-estimate}
 |H_Zy|\ge\Bigl(\frac12-6\varepsilon\Bigr)D_1D_2|y|_w,
\end{equation}
where $\kappa>0$ depends only on the degrees, the densities $c_1,c_2$,
and the orbit counts (\cite[Lemma~12]{LH22Quantum}). Reversing both
bipartitions exchanges $H_X$ with $H_Z$ and $D_i$ with $D'_i$.

\begin{proof}[Proof of \cref{thm:intro-independent-checks}]
Put $\varepsilon=1/24$ and $a_{\min}=\lceil4\gamma_0\rceil$.

\paragraph{The two graphs.}
As in the proof of \cite[Theorem~11]{LH22Quantum}, we choose the graphs in
order, since the loss of the second graph must be small compared with
the degrees of the first. Apply \cref{codingcompletion:prop:graphs} with
$t_1$, loss $\varepsilon_1=\varepsilon/t_1$, and $a_{\min}$. It gives
$u_1$, a density $c_1$, and a graph of degrees $(a,t_1a)$ with
$a\ge a_{\min}$. Reversing its bipartition gives a graph
$\Gamma_A$ with
\[
 A_0=G_m\times[u_1],\qquad A_1=G_m\times[t_1u_1],\qquad
 (D_1,D'_1)=(t_1a,a).
\]
Its adjacency map is injective, being the transpose of a full-row-rank
map. Now apply \cref{codingcompletion:prop:graphs} with $t_2$, loss
$\varepsilon_2=\varepsilon/(t_1a)$, and $a_{\min}$. It gives $u_2$, a
density $c_2$, and a graph $\Gamma_B$ with
\[
 B_0=G_m\times[t_2u_2],\qquad B_1=G_m\times[u_2],\qquad
 (D_2,D'_2)=(b,t_2b),
\]
where $b\ge a_{\min}$. Its adjacency map is surjective. For every $m$ above both thresholds
$m_0$, the group $G_m$ acts freely on both graphs. Since
$\max\{D_1,D'_1\}=t_1a$, with $\varepsilon_1t_1a=\varepsilon a$ and
$\varepsilon_2t_1a=\varepsilon$, the hypotheses
of~\eqref{codingcompletion:eq:LH-estimate} hold in both orientations.

\paragraph{Rank and rate.}
We have $H_XH_Z^T=(\delta_1\delta_0)^T=0$, and both check matrices have
full row rank by \cref{thm:LH-rank-inheritance}. Freeness gives
$|V_{ij}|=|A_i||B_j|/|G_m|$. With $M=u_1u_2|G_m|$, the blocks
$V_{00},V_{10},V_{01},V_{11}$ have sizes
\[
 t_2M,\qquad t_1t_2M,\qquad M,\qquad t_1M.
\]
The qubits are $V_{10}\sqcup V_{01}$, and the ranks of $H_X,H_Z$ are
$|V_{00}|,|V_{11}|$. Hence
\[
 N=(t_1t_2+1)M,\qquad K=N-t_2M-t_1M=(t_1-1)(t_2-1)M.
\]

\paragraph{Small-set soundness and distance.}
Let $x$ be an $X$-error, and let $y$ minimize $|\cdot|_w$ over
$x+\operatorname{im}H_X^T$. Then $y$ is locally minimal. Put $d=\min\{D_1,D_2\}$ and
$D=\max\{D_1,D_2\}$. The coset contains an element $x'$ with
$|x'|=r_X(x)$, so minimality gives
$|y|_w\le|x'|_w\le|x'|/d=r_X(x)/d$, and hence
$|y_{10}|,|y_{01}|\le D|y|_w\le(D/d)\,r_X(x)$. Recall that
$N=(t_1t_2+1)u_1u_2|G_m|$. Choose $\alpha>0$ below
$\kappa d/(D(t_1t_2+1)u_1u_2)$ and below the analogous bound for the
reversed bipartitions. Neither bound depends on $m$. If
$r_X(x)\le\alpha N$, then $|y_{10}|,|y_{01}|\le(D/d)\,\alpha N<\kappa|G_m|$,
so~\eqref{codingcompletion:eq:LH-estimate} applies to $y$ and
\[
 |H_Zx|=|H_Zy|\ge\frac14D_1D_2|y|_w
 =\frac14\bigl(D_2|y_{10}|+D_1|y_{01}|\bigr)
 \ge\frac d4|y|\ge\frac d4\,r_X(x).
\]
Since $D_1=t_1a$, $D_2=b$, and $a,b\ge a_{\min}$, we have
$d\ge a_{\min}\ge4\gamma_0$, so
$|H_Zx|\ge\gamma_0r_X(x)$. Reversing both bipartitions gives
$|H_Xz|\ge\gamma_0r_Z(z)$ in the same way, since also
$\min\{D'_1,D'_2\}\ge a_{\min}$. This
proves small-set soundness with soundness coefficient $\gamma_0$. A
nontrivial logical operator has a nontrivial $X$- or $Z$-part, say $x$,
with $H_Zx=0$ and $r_X(x)>0$. The inequality forces $r_X(x)>\alpha N$, so
the quantum minimum distance is greater than $\alpha N$.

\paragraph{Sparsity and explicitness.}
As in \cite[proof of Theorem~9]{LH22Quantum}, the $X$- and $Z$-checks
have weights $D_1+D_2$ and $D'_1+D'_2$, and the qubits in $V_{10}$ and
$V_{01}$ lie in $D'_1+D_2$ and $D_1+D'_2$ checks. All of these are at
most $w=t_1a+t_2b$.

All orbit counts, degrees, and primes are fixed before $m$ grows. A
vertex of the balanced product has coordinates
\[
 [(g,i),(h,j)]\longleftrightarrow(i,g^{-1}h,j),
\]
so its neighbors are enumerated using the neighbor algorithms of the two
graphs and a fixed number of group operations in $G_m$. This takes time
polynomial in $m=O(\log N)$, for both rows and columns of $H_X$ and $H_Z$.
\end{proof}

\section*{Acknowledgments}
We are grateful to the Israeli Interdisciplinary Center for the Theory
of Quantum Computing, at whose conference we first learned of this
problem. We also want to thank Dorit Aharonov, Alex Lubotzky, Shai Evra and Nadav Drechsler for many interesting conversations.

\section*{AI methodology}
We used ChatGPT (OpenAI) and Claude (Anthropic) extensively in this
work. ChatGPT assisted with the development of the main ideas and
proofs. This includes the strategy of using free actions of finite
$2$-groups to reduce the rank conditions to a problem on a fixed finite
gadget, and help with the rank arguments and the full-rank gadget modification, and
parts of the arithmetic construction. Both tools were used to examine
proofs for gaps, to derive soundness estimates and parameter choices,
for exploratory computations, and to draft and
revise the manuscript. Claude was used in particular to check and
simplify proofs and to restructure the presentation. The authors verified the correctness and originality of
all content, including references.

\bibliographystyle{alpha}
\bibliography{__references}

@article{BE21,
  author = {Breuckmann, Nikolas P. and Eberhardt, Jens N.},
  title = {Balanced product quantum codes},
  journal = {IEEE Transactions on Information Theory},
  number = {10},
  volume = {67},
  pages = {6653--6674},
  year = {2021},
  doi = {10.1109/TIT.2021.3097347},
  eprint = {2012.09271},
  archivePrefix = {arXiv},
  primaryClass = {quant-ph},
  url = {https://doi.org/10.1109/TIT.2021.3097347},
  note = {\href{https://doi.org/10.1109/TIT.2021.3097347}{doi:\nolinkurl{10.1109/TIT.2021.3097347}}. Preprint: \href{https://arxiv.org/abs/2012.09271}{arXiv:\nolinkurl{2012.09271}}}
}

@article{BMOR18,
  author = {Bennett, Michael A. and Martin, Greg and O'Bryant, Kevin and Rechnitzer, Andrew},
  title = {Explicit bounds for primes in arithmetic progressions},
  journal = {Illinois Journal of Mathematics},
  number = {1--4},
  volume = {62},
  pages = {427--532},
  year = {2018},
  doi = {10.1215/ijm/1552442669},
  eprint = {1802.00085},
  archivePrefix = {arXiv},
  primaryClass = {math.NT},
  url = {https://doi.org/10.1215/ijm/1552442669},
  note = {\href{https://doi.org/10.1215/ijm/1552442669}{doi:\nolinkurl{10.1215/ijm/1552442669}}. Preprint: \href{https://arxiv.org/abs/1802.00085}{arXiv:\nolinkurl{1802.00085}}}
}

@article{BSGKSV10,
  author = {Ben-Sasson, Eli and Guruswami, Venkatesan and Kaufman, Tali and Sudan, Madhu and Viderman, Michael},
  title = {Locally testable codes require redundant testers},
  journal = {SIAM Journal on Computing},
  number = {7},
  volume = {39},
  pages = {3230--3247},
  year = {2010},
  doi = {10.1137/090779875},
  url = {https://doi.org/10.1137/090779875},
  note = {\href{https://doi.org/10.1137/090779875}{doi:\nolinkurl{10.1137/090779875}}}
}

@techreport{CLRTT26,
  author = {Cohen, Itay and Leigh, Itai and Reiner, Assaf and Ta-Shma, Amnon and Tzalik, Elad},
  title = {Good quantum locally testable codes from lossless cubical complexes},
  institution = {Electronic Colloquium on Computational Complexity (ECCC)},
  type = {Report},
  number = {\href{https://eccc.weizmann.ac.il/report/2026/202/}{TR26-202}},
  year = {2026},
  eprint = {2610.00525},
  archivePrefix = {arXiv},
  note = {Preprint: \href{https://arxiv.org/abs/2610.00525}{arXiv:\nolinkurl{2610.00525}}}
}

@article{HGL25,
  author = {Hong, Yifan and Guo, Jinkang and Lucas, Andrew},
  title = {Quantum memory at nonzero temperature in a thermodynamically trivial system},
  journal = {Nature Communications},
  volume = {16},
  pages = {316},
  year = {2025},
  doi = {10.1038/s41467-024-55570-7},
  eprint = {2403.10599},
  archivePrefix = {arXiv},
  primaryClass = {quant-ph},
  url = {https://doi.org/10.1038/s41467-024-55570-7},
  note = {Article 316. \href{https://doi.org/10.1038/s41467-024-55570-7}{doi:\nolinkurl{10.1038/s41467-024-55570-7}}. Preprint: \href{https://arxiv.org/abs/2403.10599}{arXiv:\nolinkurl{2403.10599}}}
}

@inproceedings{HLMRZ25,
  author = {Hsieh, Jun-Ting and Lubotzky, Alexander and Mohanty, Sidhanth and Reiner, Assaf and Zhang, Rachel Yun},
  title = {Explicit lossless vertex expanders},
  booktitle = {2025 IEEE 66th Annual Symposium on Foundations of Computer Science (FOCS)},
  pages = {894--911},
  publisher = {IEEE},
  year = {2025},
  doi = {10.1109/FOCS63196.2025.00046},
  eprint = {2504.15087},
  archivePrefix = {arXiv},
  primaryClass = {math.CO},
  url = {https://arxiv.org/abs/2504.15087v1},
  note = {\href{https://doi.org/10.1109/FOCS63196.2025.00046}{doi:\nolinkurl{10.1109/FOCS63196.2025.00046}}. Full version: \href{https://arxiv.org/abs/2504.15087v1}{arXiv:\nolinkurl{2504.15087v1}}. Numbered statements refer to this full version}
}

@misc{LH22Quantum,
  author = {Lin, Ting-Chun and Hsieh, Min-Hsiu},
  title = {Good quantum {LDPC} codes with linear time decoder from lossless expanders},
  year = {2022},
  doi = {10.48550/arXiv.2203.03581},
  eprint = {2203.03581},
  archivePrefix = {arXiv},
  primaryClass = {quant-ph},
  url = {https://arxiv.org/abs/2203.03581v1},
  howpublished = {\href{https://arxiv.org/abs/2203.03581v1}{arXiv:\nolinkurl{2203.03581v1}}},
  note = {Version 1. Numbered statements refer to this version}
}

@article{LPS88,
  author = {Lubotzky, Alexander and Phillips, Ralph and Sarnak, Peter},
  title = {{Ramanujan} graphs},
  journal = {Combinatorica},
  number = {3},
  volume = {8},
  pages = {261--277},
  year = {1988},
  doi = {10.1007/BF02126799},
  url = {https://doi.org/10.1007/BF02126799},
  note = {\href{https://doi.org/10.1007/BF02126799}{doi:\nolinkurl{10.1007/BF02126799}}}
}

@book{Lub94,
  author = {Lubotzky, Alexander},
  title = {Discrete Groups, Expanding Graphs and Invariant Measures},
  series = {Progress in Mathematics},
  volume = {125},
  publisher = {Birkh{\"a}user},
  address = {Basel},
  year = {1994}
}

@incollection{LW93,
  author = {Lubotzky, Alexander and Weiss, Benjamin},
  title = {Groups and expanders},
  booktitle = {Expanding Graphs ({P}rinceton, {NJ}, 1992)},
  series = {DIMACS Series in Discrete Mathematics and Theoretical Computer Science},
  volume = {10},
  pages = {95--109},
  publisher = {American Mathematical Society},
  address = {Providence, RI},
  year = {1993}
}

@article{Dic22,
  author = {Dickson, Leonard E.},
  title = {Arithmetic of quaternions},
  journal = {Proceedings of the London Mathematical Society (2)},
  volume = {20},
  pages = {225--232},
  year = {1922}
}

@incollection{Kne66,
  author = {Kneser, Martin},
  title = {Strong approximation},
  booktitle = {Algebraic Groups and Discontinuous Subgroups ({B}oulder, {C}olo., 1965)},
  series = {Proceedings of Symposia in Pure Mathematics},
  volume = {9},
  pages = {187--196},
  publisher = {American Mathematical Society},
  address = {Providence, RI},
  year = {1966}
}

@book{Ser73,
  author = {Serre, Jean-Pierre},
  title = {A Course in Arithmetic},
  series = {Graduate Texts in Mathematics},
  volume = {7},
  publisher = {Springer-Verlag},
  address = {New York},
  year = {1973}
}

@article{Mar73,
  author = {Margulis, Grigorii A.},
  title = {Explicit constructions of concentrators},
  journal = {Problemy Pereda\v{c}i Informacii},
  volume = {9},
  number = {4},
  pages = {71--80},
  year = {1973},
  note = {In Russian. MR0484767}
}

@inproceedings{RV05,
  title = {Derandomized squaring of graphs},
  author = {Rozenman, Eyal and Vadhan, Salil},
  booktitle = {Approximation, Randomization and Combinatorial Optimization. Algorithms and Techniques (APPROX-RANDOM 2005)},
  series = {Lecture Notes in Computer Science},
  volume = {3624},
  pages = {436--447},
  publisher = {Springer},
  year = {2005}
}

@incollection{LS03,
  author = {Lubotzky, Alexander and Segal, Dan},
  title = {Strong approximation for linear groups},
  booktitle = {Subgroup Growth},
  series = {Progress in Mathematics},
  volume = {212},
  pages = {389--407},
  publisher = {Birkh{\"a}user},
  address = {Basel},
  year = {2003},
  doi = {10.1007/978-3-0348-8965-0_26}
}

@book{PR94,
  author = {Platonov, Vladimir P. and Rapinchuk, Andrei S.},
  title = {Algebraic Groups and Number Theory},
  series = {Pure and Applied Mathematics},
  volume = {139},
  publisher = {Academic Press},
  address = {San Diego},
  year = {1994}
}

\appendix
\section{Arithmetic background}
\label{app:lps-background}

This appendix collects the standard arithmetic inputs used in
\cref{sec:2group-cubical}. They concern the integral LPS generator
sets, the action of the LPS group on the Bruhat--Tits tree, and the
Ramanujan property of congruence quotients.

\subsection{Integral quaternions and the product structure}
\label{app:lps-integral}

We use the notation of \cref{subsec:2group-lps}. Let
$D=\Lips(\mathbb Q)$ be the Hamilton quaternion algebra over
$\mathbb Q$, with the Lipschitz order $\Lips=\Lips(\mathbb Z)$. The
algebra is definite at the real place, ramified at $2$, and split at
every odd prime. Since every $a\in A(n)$ satisfies $a\equiv1\pmod2$,
products of elements of the sets $A(n)$ again have odd real part. Each
$A(n)$ is closed under conjugation. For a prime $p\equiv1\pmod4$, the
representatives with $a_0>0$ give the usual LPS generator list
\cite[Section~2.1]{Lub94}.

\begin{lemma}[{\cite[Lemma~4.8]{HLMRZ25}}]
\label{lem:lps-integral-products}
Let the primes $p_1,\ldots,p_k$ be distinct and congruent to $1$ modulo
$4$, and put $n=p_1\cdots p_k$. Then
\[
 |A(n)|=\prod_{i=1}^k(p_i+1),
\]
and, for every prescribed ordering, multiplication induces a bijection
\begin{equation}
\label{eq:lps-integral-product}
 A(p_1)\times\cdots\times A(p_k)
 \xrightarrow{\ \sim\ } A(n).
\end{equation}
In particular, $|A(p)|=p+1$ and
$A(p_i)A(p_j)=A(p_ip_j)=A(p_j)A(p_i)$ for $i\ne j$.
\end{lemma}

\begin{proof}
We recall the argument. Its factorization input is Dickson's
factorization theorem for odd quaternions \cite[Theorem~8]{Dic22}.

\emph{Counting.}
For odd squarefree $n$, Jacobi's four-square formula gives
\[
 \bigl|\{a\in\Lips:N(a)=n\}\bigr|
 =8\sum_{d\mid n}d
 =8\prod_{i=1}^k(p_i+1).
\]
Coordinate permutations show that exactly one quarter of these solutions
have their unique odd coordinate in the real position. Quotienting
that set by the fixed-point-free involution $a\mapsto-a$ divides its
size by two. This proves the displayed formula for $|A(n)|$.

\emph{Factorization and normalization.}
Let $\gamma\in\Lips$ represent an element of $A(n)$. Since $n$ is
squarefree, $\gamma$ is primitive. For a fixed ordering of
$p_1,\ldots,p_k$, the cited factorization theorem gives
\[
 \gamma=\alpha_1\cdots\alpha_k,\qquad N(\alpha_i)=p_i,
\]
uniquely up to unit migration. The units are
$\{\pm1,\pm\qi,\pm\qj,\pm\qk\}$, and the migration has the form
\[
 \alpha_i\longmapsto u_{i-1}^{-1}\alpha_i u_i,
 \qquad u_0=u_k=1.
\]
Right multiplication by $\qi,\qj,\qk$ moves the unique odd coordinate to
the other three positions. Choose $u_1,\ldots,u_{k-1}$ successively so that
the first $k-1$ factors have odd real part. Each choice is unique up to
sign. Those factors are all congruent to $1$ modulo $2$, and so is
$\gamma$. Therefore the last factor is also congruent to $1$ modulo $2$.
This gives a factorization by elements of the prescribed $A(p_i)$, unique
modulo the signs already quotiented out. It proves
\eqref{eq:lps-integral-product}.
\end{proof}

\subsection{The LPS group and the tree}
\label{app:lps-group}

Fix a prime $p\equiv1\pmod4$. For a commutative ring $R$, write
\[
 G'(R)=\Lips(R)^\times/R^\times
\]
for the group of units of $\Lips(R)$ modulo scalar units. The units of
$\Lips(\mathbb Z[1/p])$ are the elements whose norm is a power of $p$, and
the scalar units are $\pm p^z$. For the Hamilton quaternions,
$G'(\mathbb Z[1/p])$ is the group denoted $\Gamma$ in
\cite[Section~7.0]{Lub94}. For a positive integer $N$ prime to $p$, its
principal congruence subgroup of level $N$ is
\begin{equation}
\label{eq:lps-congruence-subgroup}
 \Gamma_p(N)
 =\ker\Bigl(G'\bigl(\mathbb Z[1/p]\bigr)\longrightarrow
   G'\bigl(\mathbb Z[1/p]/N\mathbb Z[1/p]\bigr)\Bigr),
\end{equation}
where $\mathbb Z[1/p]/N\mathbb Z[1/p]\cong\mathbb Z/N\mathbb Z$. Thus
$\Gamma_p(N)$ consists of the classes $[a]$ with
$a\equiv c\pmod N$ for some scalar $c$ invertible
modulo $N$. Being a kernel, it is normal and has finite index.

Let
\[
 \Lambda_p
 =\bigl\langle[a]:a\in A(p)\bigr\rangle
 \subseteq G'\bigl(\mathbb Z[1/p]\bigr).
\]
Since $a\bar a=p$ for $a\in A(p)$, one has $[\bar a]=[a]^{-1}$.
Choose a splitting $D\otimes\mathbb Q_p\cong M_2(\mathbb Q_p)$.
Let $T_{p+1}$ be the $(p+1)$-regular Bruhat--Tits tree, whose vertex set is
\[
 \operatorname{PGL}_2(\mathbb Q_p)/
 \operatorname{PGL}_2(\mathbb Z_p).
\]
The splitting gives $G'(\mathbb Z[1/p])$ an action on $T_{p+1}$.

\begin{theorem}[LPS tree realization]
\label{thm:lps-tree}
The following standard properties hold
\cite{LPS88}, \cite[Corollary~2.1.11, Lemma~7.4.1, and p.~95]{Lub94}.
\begin{enumerate}
\item[(i)] $\Lambda_p=\Gamma_p(2)$. Equivalently, $\Lambda_p$ consists
exactly of the classes having a representative
$a\in\Lips(\mathbb Z[1/p])$ with $a\equiv1\pmod2$ and $N(a)$ a power of
$p$.
\item[(ii)] $\Lambda_p$ is free on $(p+1)/2$ generators. The full
symmetric generating set is $A(p)$, with conjugate pairs representing
inverse generators.
\item[(iii)] $\Lambda_p$ acts simply transitively on the vertices of
$T_{p+1}$, and
\[
 T_{p+1}\cong\operatorname{Cay}(\Lambda_p;A(p)).
\]
\end{enumerate}
\end{theorem}

Now let $N$ be even and prime to $p$. Then
$\Gamma_p(N)\subseteq\Gamma_p(2)=\Lambda_p$, and $\Gamma_p(N)$ is the
kernel of the restriction to $\Lambda_p$ of projective reduction modulo
$N$. Write this restriction as
$\pi_N\colon\Lambda_p\to G'(\mathbb Z/N\mathbb Z)$. For $N=2^m$, it is the
map $\pi_m$ of \cref{sec:2group-cubical}, whose image lies in
$\widehat G_m\subseteq G'(\mathbb Z/2^m\mathbb Z)$. Quotienting the Cayley
graph of \cref{thm:lps-tree}(iii) by the normal subgroup
$\Gamma_p(N)$ gives the Cayley graph of the quotient group:
\begin{equation}
\label{eq:lps-congruence-cayley}
 \Gamma_p(N)\backslash T_{p+1}
 \cong\operatorname{Cay}
       \bigl(\pi_N(\Lambda_p);\pi_N(A(p))\bigr).
\end{equation}
At this stage the generating list is counted with multiplicity: collisions
or identity images have not yet been excluded. In
\cref{sec:2group-cubical}, \cref{lem:2group-nondegeneracy}
excludes them at the levels used there.

The parity of word length defines an index-two subgroup
$\Lambda_p^+\subseteq\Lambda_p$. This is well defined either from
freeness or from the bipartition of $T_{p+1}$.

\subsection{The Ramanujan property of congruence quotients}
\label{app:lps-ramanujan}

We use the following theorem from \cite[Chapter~7]{Lub94}. It extends the
Ramanujan property of the LPS graphs to the congruence quotients of every
level prime to $p$, without identifying their vertex groups. We state it
for the Hamilton quaternions and the group $G'(\mathbb Z[1/p])$ above.

\begin{theorem}[{\cite[Theorems~7.3.1 and~7.3.12]{Lub94}}]
\label{thm:lps-congruence-ramanujan}
Let $p\equiv1\pmod4$ be prime and let $N$ be a positive integer prime
to $p$. Then
\[
 X^{p,N}=\Gamma_p(N)\backslash T_{p+1}
\]
is a finite $(p+1)$-regular Ramanujan graph: every adjacency eigenvalue
$\lambda$ satisfies
\begin{equation}
\label{eq:lps-black-box-spectrum}
 \lambda=\pm(p+1)\qquad\text{or}\qquad |\lambda|\le2\sqrt p.
\end{equation}
Moreover:
\begin{enumerate}
\item[(i)] If $p$ is a quadratic residue modulo $N$, then $X^{p,N}$ is
non-bipartite and has girth at least $2\log_pN-2\log_p2$.
\item[(ii)] If $p$ is not a quadratic residue modulo $N$, then $X^{p,N}$
is bipartite and has girth at least $4\log_pN-4\log_p2$.
\end{enumerate}
\end{theorem}

Parts~(i) and~(ii) are Parts~II and~I of \cite[Theorem~7.3.12]{Lub94},
respectively. The deep spectral input is Deligne's theorem
\cite[Theorem~6.1.2]{Lub94}, the Ramanujan--Petersson conjecture in
representation-theoretic form. It is transferred to the quaternion algebra
by the Jacquet--Langlands correspondence
\cite[Section~6.2, Theorem~6.2.1 and Corollary~6.2.2]{Lub94}, and
\cite[Theorem~7.1.1]{Lub94} records the resulting statement for the
congruence subgroups. We use the theorem as a black box, and no
automorphic argument is repeated here.

\end{document}